\documentclass[%
 reprint,
 amsmath,amssymb,
 aps,
]{revtex4-2}

\usepackage{graphicx}
\usepackage{dcolumn}
\usepackage{bm}
\usepackage{xcolor}
\usepackage{appendix}

\begin{document}

\preprint{APS/123-QED}

\title{Training a multilayer dynamical spintronic network with standard \\ machine learning tools to perform time series classification}

\author{Erwan Plouet}
\author{Dédalo Sanz-Hernández}
\author{Aymeric Vecchiola}
\author{Julie Grollier}
\author{Frank Mizrahi}%
 \email{frank.mizrahi@cnrs-thales.fr}
\affiliation{Laboratoire Albert Fert, CNRS, Thales, Université Paris-Saclay, 91767 Palaiseau, France}%

\date{\today}

\begin{abstract}
The ability to process time-series at low energy cost is critical for many applications. Recurrent neural network, which can perform such tasks, are computationally expensive when implementing in software on conventional computers. Here we propose to implement a recurrent neural network in hardware using spintronic oscillators as dynamical neurons. Using numerical simulations, we build a multi-layer network and demonstrate that we can use backpropagation through time (BPTT) and standard machine learning tools to train this network. Leveraging the transient dynamics of the spintronic oscillators, we solve the sequential digits classification task with $89.83\pm2.91~\%$ accuracy, as good as the equivalent software network. We devise guidelines on how to choose the time constant of the oscillators as well as hyper-parameters of the network to adapt to different input time scales.
\end{abstract}

\maketitle


\section{\label{Intro}Introduction}

The ability to process time-series (classification, prediction, generation etc.) is important for many applications from smart sensors in industrial maintenance to personal assistants and medical devices. 
Using the dynamics of a physical system, leveraging its non-linearity and memory for such processing has been widely explored with the development of recurrent neural networks, both from a purely mathematical perspective \cite{pineda1987generalization, lecun1992theoretical, hermans2013training, hasani2021liquid} as well as from a brain-inspired perspective with spiking recurrent neural networks \cite{eshraghian2023training, taherkhani2020review, tavanaei2019deep, yin2021accurate}. Chen et al. have shown that neural networks based on ordinary differential equations (Neural-ODEs) can be seen as residual neural networks where the time dimension acts as depth thus providing computing power \cite{NODE}. These findings highlight the potential of dynamical systems to implement deep neural networks. While implementing a recurrent neural network in software on a conventional computer requires computing the evolution of each neuron step by step, a physical network would naturally perform this computation. 
As there is an increasing demand for learning and processing on the edge, with strong footprint and energy cost constraints, building novel hardware that directly implements dynamic recurrent neural networks is an attractive path. This motivation has led to the realisation and training of recurrent neural networks with a wide variety of dynamic systems: analogue CMOS \cite{pehle2022brainscales}, photonic systems \cite{bueno2018reinforcement}, acoustic resonators \cite{qu2022resonance}, mechanical oscillators \cite{coulombe2017computing} and wave systems \cite{hughes2019wave} to cite a few.

Spintronic oscillators are promising building blocks for the hardware implementation of neural networks, due to their non-linear high-speed dynamics as well as potential for miniaturization and CMOS integration \cite{finocchio2021promise, grollier2020neuromorphic}. Time-series processing using the dynamics of spintronic oscillators has been repeatedly demonstrated experimentally in the context of reservoir computing, thus validating they have the required non-linearity and memory properties  \cite{tanaka2019recent, torrejon2017neuromorphic, markovic2019reservoir, tsunegi2019physical, kanao2019reservoir, shreya2023granular, tsunegi2018evaluation}. However, reservoir computing is limited because only the output classifier is trained and the dynamics is fixed. Training of spintronic oscillators for time-series processing has been limited to single layer networks \cite{romera2018vowel, zahedinejad2020two}. 
Ross et al. have experimentally demonstrated a multilayer network of spintronic oscillators, but with a feedforward architecture dedicated to static tasks \cite{Andrewmulti}. Rodrigues et al. have shown by numerical simulations how to train the transient dynamics of a single layer network of oscillators with optimal control theory, on a static task \cite{DynFinnochio}.

Here we simulate and train a multi-layer network of spintronic oscillators as neurons, using standard machine learning tools. We leverage the transient dynamics of the oscillators to perform time-series classification of the sequential digits dataset. First, we describe the dynamic spintronic neuron model, the architecture of the network and how to cascade neuron layers. Second, we describe how to train the network with PyTorch and backpropagation through time (BPTT) and demonstrate $89.83\pm2.91~\%$ accuracy on sequential digits, as good as the equivalent software network. Third, we show that the spintronic network can be trained on input timescales over a 5-fold range, centered around a value depending on device parameters. Finally, we derive guidelines to find these device parameters. In particular, the neuron relaxation time must be larger than the input time scale and the cumulative drive must be around one. 

\section{\label{network idea}Multilayer network of spintronic dynamical neurons}

\begin{figure*}
\onecolumngrid
    \centering
    \includegraphics[width=\textwidth]{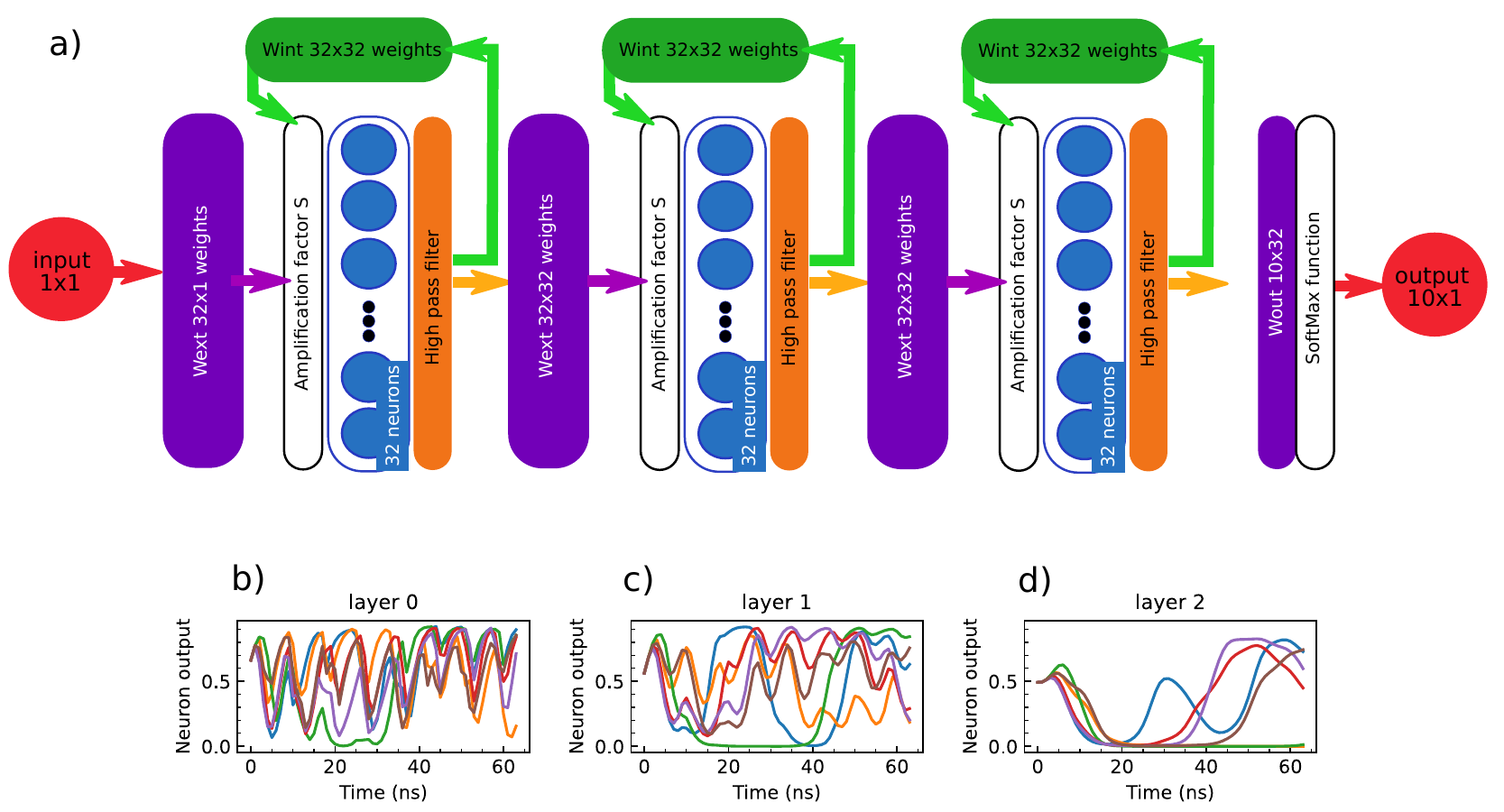}

    \caption{ Architecture of the network. (a) Schematic of the network architecture with three layers of 32 neurons each. The neurons are represented by blue circles, the interlayer connections ($W_{ext}$) by purple boxes, the intralayer connections ($W_{int}$) by green boxes, the amplification factors and SoftMax by white boxes, and the high pass filters by orange boxes. The dimensions of the connections are indicated in the corresponding boxes. At each time step the input is of size one. The output is of size 10. For simplicity, we do not represent the fixed and trainable biases on this figure. (b-c-d) Normalized output RF power versus time, emitted respectively by the neuron 0 from layer 0, the neuron 0 from layer 1 and the neuron 1 from layer 2 . The colors (blue, orange, green, red, purple, brown) correspond to different inputs from different classes (respectively 5,6,9,3,3,8).}
    \label{fig:network}
\end{figure*}

Spintronic oscillators exhibit magnetization self-sustained oscillations when a dc current drive overcomes the magnetic damping. Magnetoresistive effects translate the magnetization oscillations into voltage oscillations that can be fed as input to other devices.  In this work, we use the amplitude of these oscillations as internal variable of the spintronic neuron. When submitted to a drive, the oscillation amplitude undergoes a transient regime which provides memory to the device. We use the auto-oscillator model \cite{slavin_nonlinear_2009} to describe the dynamics of the neurons. We choose this model because it has been repeatedly verified by experiments, in particular in the context of dynamical neurons \cite{DynFinnochio} . Using this model rather than directly solving the Landau–Lifshitz–Gilbert (LLG) equation or performing micromagnetic simulations has two advantages. First, it reduces simulation length, making it possible to train multilayer neural networks with hyperparameter search, which is critical for physical neural networks, as will be detailed in Sections III and IV. Second, it provides a general model, which can be applied widely to different types of spintronic oscillators. Slavin's model describes the dynamics of the magnetization dimensionless normalized amplitude $x$:

\begin{equation}
    \frac{dx}{dt}=-2(\gamma(1+Qx)-\sigma I(t)(1-x))x
\end{equation}

Where $\gamma=\alpha\times\omega$ is a damping term with $\alpha$ the magnetic Gilbert damping and $\omega$ the frequency of the device. $Q$ is the non-linear damping, which can take various values depending on the exact materials, drives and devices used for the neurons. We set $Q = 0$ in order to show the intrinsic non-linearity of spintronic oscillator sufficient for time-series classification. A non-zero $Q$ provides additional non-linearity and computing power. $\sigma$ is a material parameter which we set to one, as it simply rescales $I(t)$. In this work, we simplify the magnetization dynamics into:

\begin{equation}
    \frac{dx}{dt}=-\gamma x +I(t)x(1-x)
\end{equation}

$I(t)$ is the time dependent drive, a dimensionless variable proportional to the dc current injected into the device. The damping $\gamma$ pulls the output to zero and $x(x-1)$ is the term that acts as a natural bounding of the output between 0 and 1. The oscillator exhibits non-linearity as well as memory, making it suitable for processing time series. Furthermore, it follows the shape $x(t+1)=g(x(t), Input(t))$, which is the conventional shape of a recurrent neuron. Here, the hidden state variable of the neuron $x$ is the oscillation power.

We consider a network, depicted in Figure 1, composed of successive layers of dynamical neurons, represented by blue circles. The neurons are connected by interlayer connections ($W_{ext}$, in purple) and intralayer connections ($W_{int}$, in green). In the scope of this paper, we consider standard linear connections, performing weighted sums of the RF power outputs of the neurons.

In order to exploit the whole dynamical range of the neurons, their output should be superior to zero (no output) and inferior to one (saturation). This requires the outputs of the neurons in the absence of external drive to be half of their maximum output power. In principle, the network could achieve this by learning the correct values for the trainable bias applied to each neuron. However, it is more efficient to apply a fixed bias $b_{fixed} $ on top the trainable biases $b_i^{n}$ to each neuron. The value of the fixed bias is twice the damping $b_{fixed} = 2\gamma$.  Furthermore, we use a high-pass filter of cut frequency $f_{cut}=1$ MHz as well as a tunable amplification factor set to $S = 0.5$ between each layer to keep the drives centered around zero. 

The dynamics of the power $x_{i}^{n+1}$ of neuron i from layer $n+1$ is given by:

\begin{equation}
    \frac{dx_{i}^{n+1}}{dt}=-\gamma x_{i}^{n+1} +I(t)x_{i}^{n+1}(1-x_{i}^{n+1})
\end{equation}

With the drive:
\begin{equation}
    I(t)=S\times(W_{i,j}^{n,ext}y_j^n+W_{i,j}^{n+1,int}y_j^n + b_i^{n+1})+b_{fixed}
\end{equation}

Where $y_j^n$ is the output of neuron $i$ from layer $n$ after application of the high pass filter and follows:
\begin{equation}
        \frac{dy_{i}^{n}}{dt}=-2\pi f_{cut}\times y_{i}^{n}+\frac{dx_{i}^{n}}{dt}
\end{equation}

Figures 1(b-c-d) illustrate the dynamics of the neurons for layers 0, 1 and 2 respectively. Each color is one time-varying input applied to the neuron. We observe varied dynamics around 0.5 (i.e. half of the maximum amplitude of the oscillations). Each layer acts an integrator, slowing down information as it flows into the network.

\section{\label{time series}Processing of a time series task}

\begin{figure*}
\onecolumngrid
    \centering
    \includegraphics[width=\textwidth]{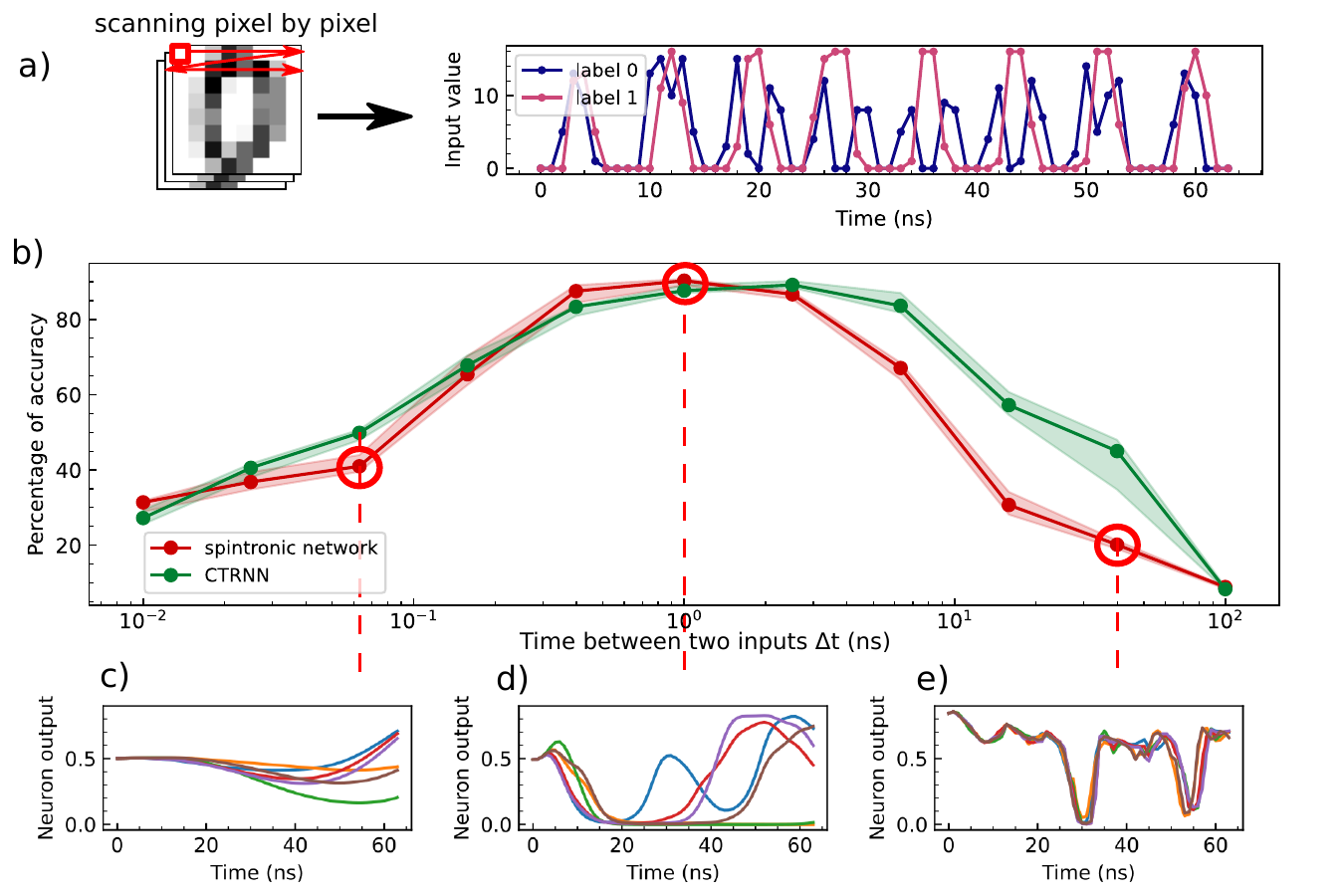}

    \caption{Performance on the sequential digits classification task. (a) Schematic of the task: the images are scanned row by row and the pixel intensity in used as input to the network. The violet and pink curves represent the input values versus time for an image of label 0 and label 1 respectively. (b) Accuracy of the network versus the time interval between two input points. The red curve corresponds to the simulated spintronic network while the green curve corresponds to a standard continuous-time recurrent neural network (CTRNN). Each accuracy point corresponds to the median accuracy on a series of 10 random initialisations, the shaded areas correspond to the range between the $75^{th}$ and $25^{th}$ percentiles. (c-d-e) Normalized output versus time for respectively the neuron 3, the neuron 1 and the neuron 2, of layer 2. The colors (blue, orange, green, red, purple, brown) correspond to different inputs from different classes (respectively 5,6,9,3,3,8). Each of the three panels corresponds to a different time interval between input points, shown by red circles and red dashed lines in panel (b). The neuron dynamics is respectively too slow, adapted, and too fast for the input time scale.}
    \label{fig:time_scan}
\end{figure*}

The time-series classification task we chose to evaluate the spintronic network is sequential digits \cite{scikit-learn}. The dataset is composed of 1797 grayscale 8-by-8-pixel images of handwritten digits and the goal is to identify the digit (labels 0 to 9). We split the dataset 50/50 between train and test. The images are presented pixel by pixel to the network, as time series of 64 input points each, as depicted in Figure 2(a). Here the time interval between input points is 1 ns. In consequence, the input of the network is of size 1 and the output of size 10. We chose this task because the digits dataset is well known and freely available, which will simplify future replication studies and benchmarking. We present the pixels sequentially in order to evaluate ability of the network to process time-series, which makes the task harder than the standard static image classification task.
The network is simulated using the PyTorch library as well as the differential equation (2). After the final time step of the series, we compute the log-likelihood loss of the output. We reset the internal state of the neurons after processing each image as there is no time-correlation between successive images. We update the trainable parameters (weights and biases) using BPTT and batches of 120 images \cite{robinson1987utilityBPTT, werbos1988generalization, mozer2013focused}. 
We clip the amplitude of each gradient element to 1 to prevent gradient explosion. We set the connectivity density at 0.5 both for the interlayer and intralayer connections ($d_{inter} = 0.5$ and $d_{intra} = 0.5$). The effect of connection density is studied in Section IV.  Figure \ref{fig:network}(d) shows the behavior of neuron 1 of the output neuron. We observe different evolutions for the different classes, i.e. colors, illustrating the separation property of the network, as well as similar evolutions for the purple and red curves which correspond to the same class. We use a hyperparameter optimization using the library Optuna \cite{optuna_2019}, in order to maximize the accuracy, and obtain a relaxation time of $\tau = 14.12~ns$ for the neurons. We use a learning rate decay method $lr = \frac{lr_0}{\frac{n_{epoch}}{lr_{decay}} + 1}$ with $lr_0=0.149$ and $lr_{decay}=8.077$, as well as the Adam optimizer \cite{kingma2017adammethodstochasticoptimization}.
We achieve an accuracy of $89.83\pm2.91~\%$.  The relaxation time corresponds to $\tau = 1/\gamma = 1/\alpha \omega$. Typical magnetic damping values in spintronic oscillators are around $\alpha = 0.01$. In consequence, the optimal frequency for the oscillators would be around $\omega  \approx 7.1 $ GHz, which is achieved with current devices \cite{choi2022voltage}. 

We benchmark this result by comparing the simulated spintronic network to a standard software-based standard continuous-time recurrent neural network (CTRNN) \cite{funahashi1993approximation}. We consider a CTRNN with the same architecture (3 layers of 32 neurons, inter and intralayer connections with density of 0.5 each). 

The hidden state $x_i^{n+1}$ of the $i$-th neuron of layer $n+1$ is driven by:

\begin{equation}
    \frac{dx_{i}^{n+1}}{dt}=-\gamma x_{i}^{n+1} +I(t)
\end{equation}

With the drive:
\begin{equation}
    I(t)=S\times(W_{i,j}^{n,ext}y_j^n+W_{i,j}^{n+1,int}y_j^n + b_i^{n+1})
\end{equation}

Where $y_j^n$ is the output of the $j$-th neuron of layer $n$ with application of a non-linear activation function:
\begin{equation}
        y_j^n=tanh(x_j^n)
\end{equation}

We see here that the spintronic network is a natural implementation of a CTRNN by a physical system. The main difference is that the internal state of the spintronic neuron is intrinsically bounded, while it is not for the software neurons, which rely on their activation function to bound their outputs. This constraint is common to all physical implementations of neurons. We observe that the CTRNN achieves a top accuracy of $89.00\pm3.48\%$. The uncertainty corresponds to the standard deviation over ten runs, each with a different random initialisation of the trainable parameters.
We conclude that the spintronic network performs as well as a standard CTRNN, despite not being isomorphic to the CTRNN equations, and in particular having the strong constraint of having neurons with bound internal state.

\section{\label{adaptation}Adaptation of the network to different time scales}

\begin{figure*}
\onecolumngrid
    \centering
    \includegraphics[width=\textwidth]{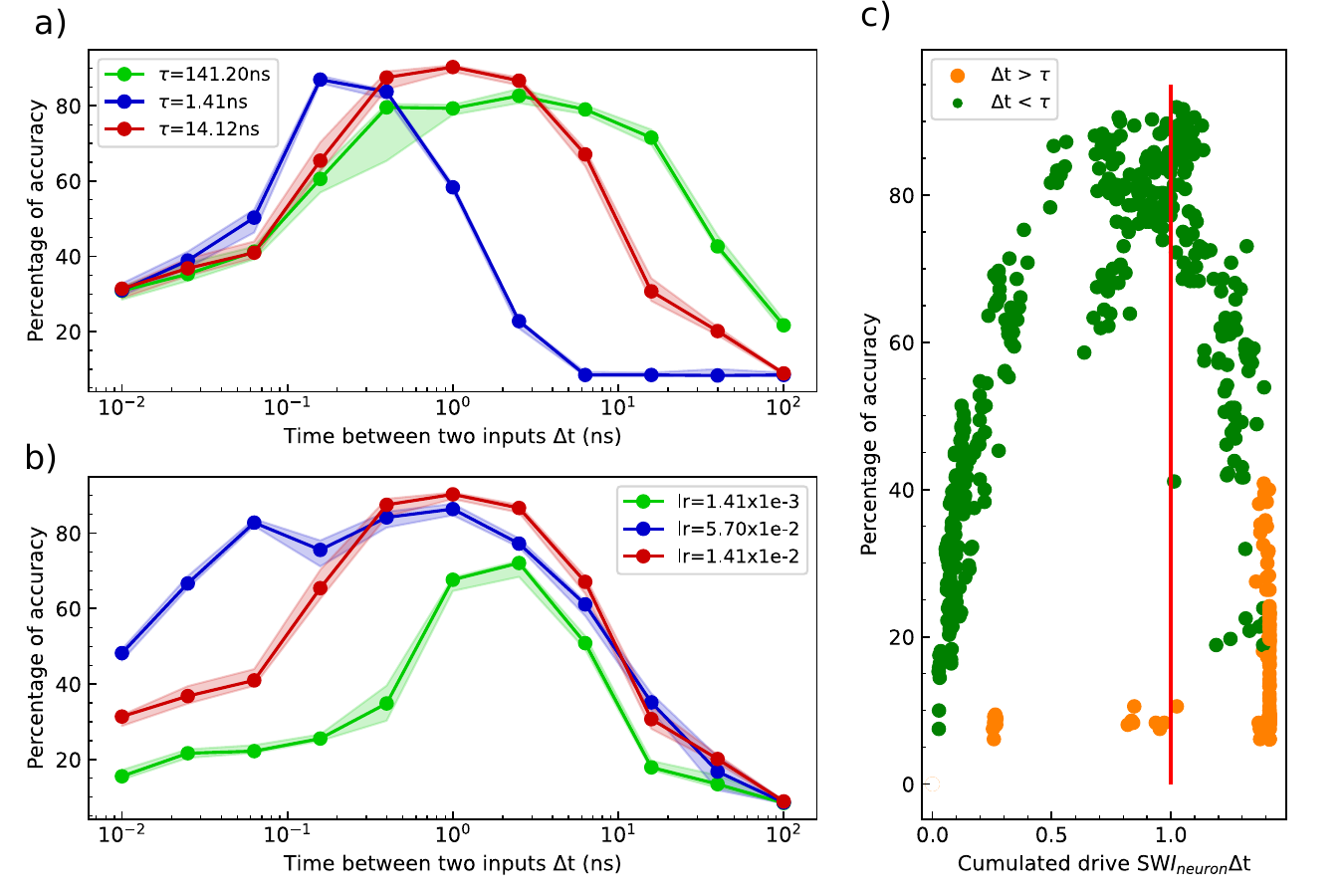}

    \caption{Time scale adaptation.  (a) Accuracy versus the time between two input points $\Delta t$ for three neuron relaxation times $\tau$ (colors). (b) Accuracy versus the time between two input points $\Delta t$ for three learning rates (colors). For (a) and (b), each point corresponds to the median accuracy on a series of 10 random initialisations, the shaded areas correspond to the range between the $75^{th}$ and $25^{th}$ percentiles. (c) Accuracy versus the cumulative drive, for a wide range of parameters. Each point corresponds to the median accuracy on a series of 10 random initialisations. The points in green (resp. orange) have a neuron relaxation time $\tau$ larger (resp. smaller) than the input time scale $\Delta t$. The vertical red line marks where the cumulative drive is equal to 1.}
    \label{fig:lr_tau}
\end{figure*}

When processing time series, a critical feature is the time scale of the input. Because the memory of the network comes from the dynamics of the physical system, the system needs to be designed to match the time scales relevant to the task. We derive design rules by studying how the network behaves at different input time scales.

We set the hyper-parameters of the network to the values found by optimization for 1 ns between input points, then train the network on series with different input time scales. Figure 2(b) shows how the accuracy depends on the time interval between two input points. 
We observe that the maximum accuracy is achieved for 1 ns, which is expected as this is the optimization point. This maximal accuracy is maintained over a wide window of time scales, about a 5-fold factor, from 0.4 to 2.5 ns. When the time scale is well below or well above 1 ns, the accuracy is significantly degraded. 
Here again, the spintronic network performs as well as the CTRNN.
Panels (c-d-e) of Figure 2 show the evolution of a neuron from layer 2 (i.e. output layer) for different inputs (colors) of different classes, in three regimes. In panel (d), the dynamics of the network is optimized for the input time scale. We observe that the different inputs produce rich varied dynamics, enabling separation of the inputs into classes. In panel (c), the input is faster and the neuron is too slow to follow its changes. We observe that the dynamics of the input is completely lost in the smoothed trajectories of the neuron, preventing accurate classification. In panel (e), the input is slower and the neuron is too fast. The memory of the network is too short to remember enough input points to achieve accurate classification.

Building on these results, we devise guidelines to design networks adapted to learn a task with a given time scale. The dynamics of the spintronic neuron relies on a competition between the current drive and the damping. The damping pulls the output amplitude to zero while the current drive pushes the neuron away from this rest position.

Figure 3(a) shows the accuracy versus the time between two input points, for three different relaxation times. We observe that the optimal time scale shifts with the relaxation time. The dependence of the accuracy is always a bell-shaped curve, centered around the optimal time. In particular, slower inputs require longer relaxation times so that the network has a long enough memory to remember the input points necessary for classification. 

Figure 3b shows the accuracy versus the time between two input points for three different learning rates. We observe that here as well, learning requires a matching between learning rate and input time scale. Furthermore, we observe that faster inputs require higher learning rates. Indeed, the neuron receives a cumulative drive $I \times \Delta t$ for each input point. The strength of this drive depends on the weights and the scaling factors:
\begin{equation}
        I=S\times \langle |W| \rangle\times \sqrt{N_{neurons}}
\end{equation}

Where here $N_{neurons}=d_{inter}\times32+d_{intra}\times32=32$ is the fan-in for each neuron and $\langle |W| \rangle$ is the mean of the absolute value of the weight elements.
When the input is fast (low $\Delta t$), the network needs to reach higher weights to provide a cumulative drive high enough to counter-balance the damping. Higher learning rates allow the weights to reach higher values.

We identify two conditions. First, the relaxation time of the neurons should be larger than the time between two input points $\Delta t < \tau$. Second, the cumulative drive divided by the relaxation time should be roughly equal to the damping:
\begin{equation}
        I\times \frac{\Delta t}{\tau} \approx \gamma
\end{equation}
As the damping is the inverse of the relaxation time, this simplifies as:
\begin{equation}
        I\times \Delta t \approx 1
\end{equation}
Figure 3c shows the accuracy versus the cumulative drive for a broad range of conditions. We observe that the best accuracies are indeed obtained when the cumulative drive is around 1, provided that the input scale is larger than the relaxation time (green dots). 
Furthermore, we recommend to set the learning rate in the range of one hundredth of the optimal weights values $\frac{1}{\Delta t\times S\times \sqrt{N_{neurons}}}$.

Note that the cumulated drive must take device and material parameters into account. For instance, if $\sigma$ is different from 1, the factor $S$ should be rescaled. We observe that the design rule regarding the cumulative drive does not require fine tuning. In Figure 2b as well as Figure 3, we observe that the accuracy is not significantly degraded when the input time scale varies within a wide window, typically 5-fold centered around the optimal value, which is promising for practical realizations. 
In this work we have provided an example of how to perform hyperparameter tuning for a simplified set of device and material parameters. For designing and training a physical network, it will be necessary to perform hyperparameter tuning with the actual device and material parameters.

\section{\label{archi}Architecture design}

We evaluate the impact of the connection density. Figure 4 shows the accuracy of the task versus the density of connections. Here the network is randomly sparsified before training.  We observe that the intralayer connectivity (red curve) can be reduced to $10 \%$ without losing accuracy. When removing the intralayer connections altogether, the accuracy is degraded but remains high. The interlayer connectivity (green curve) can be reduced to $20 \%$ without losing accuracy (87.22 \%). We also observe that the performance exhibits higher run-to-run variability at higher densities of connection. This is due to the fact that higher connectivity increases the risk of neuron saturation and thus failure of learning.  The global connectivity (both interlayer and intralayer) is optimal around $50 \%$ (90.28\% accuracy) and the accuracy is still very high down to 30\% density (85.42\%). At 20\% global connectivity, the accuracy is still high (77.78\%) but has significantly dropped. Removing connections helps to prevent overfitting, as shown in the Appendix. Removing the interlayer connections altogether prevents any learning, as expected, as the information cannot flow from input to output. 

\begin{figure}
    \centering
    \includegraphics[width=0.5\textwidth]{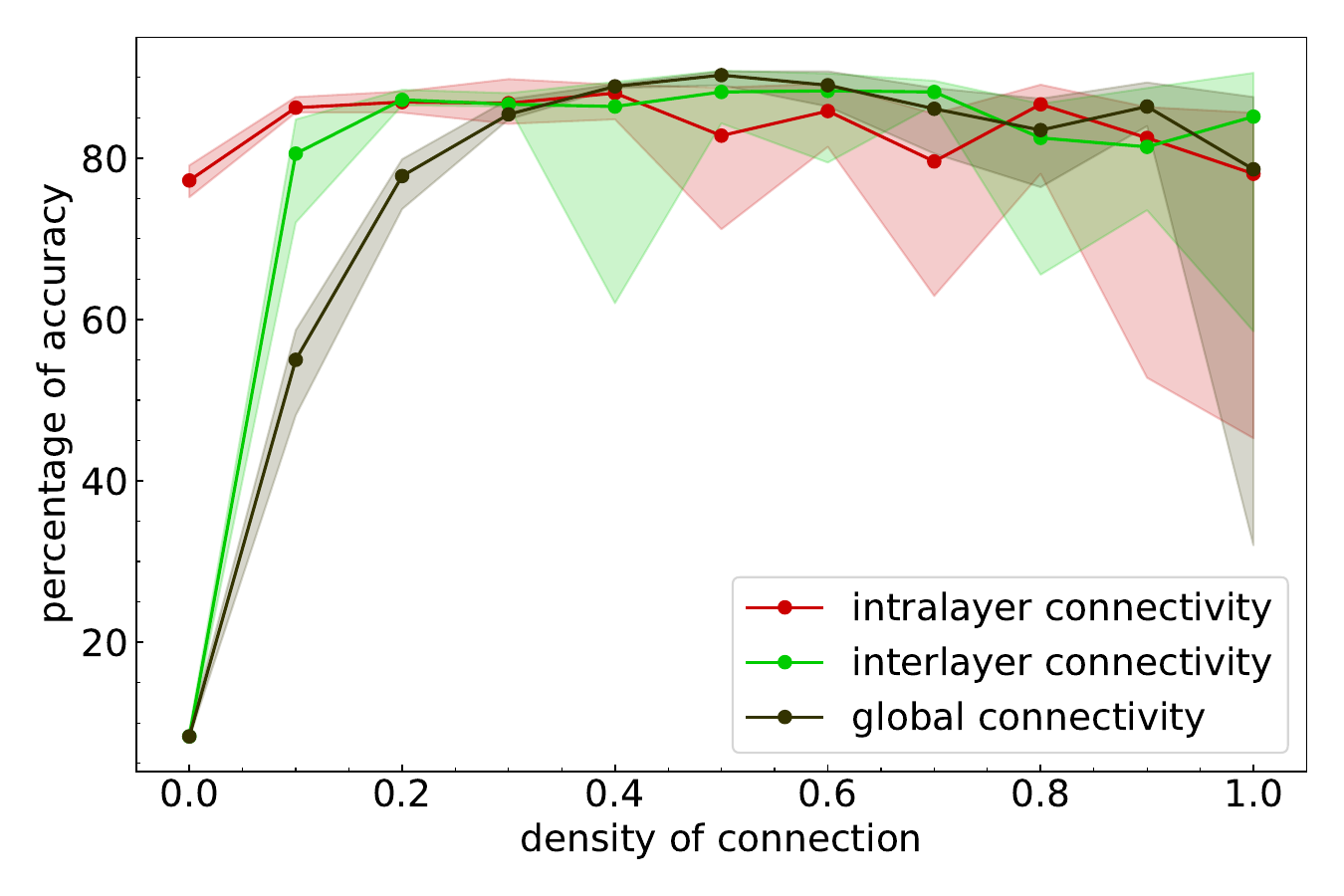}

    \caption{Effect of the density of connections. The red curve shows the accuracy versus the density of the intralayer connections ($W_{int}$), for an interlayer density of 1. The green curve shows the accuracy versus the density of the interlayer connections ($W_{ext}$), for an intralayer density of 1. The black curve shows the accuracy versus the density of all connections. Each point corresponds to the median accuracy on a series of 10 random initialisations, the shaded areas correspond to the range between the $75^{th}$ and $25^{th}$ percentiles.}
    \label{fig:density}
\end{figure}

To build such a network, we need physical implementations of the synaptic layer. These synapses need to respond fast enough to preserve the neuron dynamics. Achieving the large number of connections required for implementing state of the art artificial neural networks in hardware is challenging. However, there has been significant recent progress regarding spintronic connections. In particular, several teams have demonstrated of crossbars of magnetic tunnel junctions monolithically co-integrated with standard CMOS \cite{borders2024measurement, jung2022crossbar}. Alternatively, Ross et al. have proposed to use spintronic resonators to convert the oscillator outputs into DC signals, which provides a simpler architecture than direct coupling of oscillators \cite{Andrewmulti}.

Several technologies have been proposed to implement oscillator neurons. For instance, a CMOS oscillator at 8 GHz can be implemented with a circuit about 10 µm wide and consumes 200 µW \cite{nikonov2020convolution}. Resistive switching oscillators have a smaller footprint (down to few nm) but their frequencies are sub-GHz and their endurance is limited compared to spintronic devices \cite{nath2023harnessing}. Photonic oscillators can reach hundred GHz frequencies but require micron-sized devices. Spintronic oscillators combine small size (down to few nm) \cite{feldmann2019all}, high frequencies (above 10 GHz), low power consumption (below 10 µW) and high endurance (more than $10^{12}$ cycles). 


In the network described here, there are 96 neurons and 2896 synapses. If the image is presented for about 100 ns, according to the estimations in \cite{Andrewmulti}, each neuron and each synapse would consume 100 fJ and 10 fJ respectively, leading to a total consumption of about 40 pJ. These estimations take into account the  tunable amplifiers between layers.

In this work, we have shown the importance of hyperparameter tuning. We propose the following methodology for designing and training future hardware implementations. A first hyperparameter search is performed with software simulations for various tasks of the targeted applications, with detailed models of the specific spintronic devices to be used as neurons and synapses and taking technology constraints into account. From this search values for the oscillators frequency, fixed bias and amplifier factor are found, in order to design and build the hardware network. Oscillators frequency is set by material and geometry parameters, but can be tuned post-fabrication over wide ranges, as shown in \cite{choi2022voltage} for instance. Tunable biases, filters and amplifiers can be made with standard analogue CMOS circuits. While these are bulkier than nanodevices, only one per layer is required. 
Once fabricated, the hardware network can be trained using either off-chip training or chip-in-the-loop training. Chip-in-the-loop training is a method where the forward pass is computed by the physical network, and the backward pass is computed with a model on a computer, using the loss from the physical network. This method has the advantage of been robust to mismatch between the model and the physical network up to about 20 \% \cite{wright2022deep}. However, it requires to train separately each instance of the physical network. Furthermore, the hyperparameter tuning will be cumbersome to implement. Off-chip training makes it possible to train the network model on a computer once and then upload the trained weights on each instance of the physical network. This method enables a thorough hyperparameter tuning, as well as a single training for all nominally identical networks. Recent advances, using measurement-driven methods \cite{rasch2023hardware, borders2024measurement} as well as loss sharpness analysis \cite{xu2024perfecting} have made off-chip training robust enough to model mismatch to train physical networks. It is likely that a combination of chip-in-the-loop and off-chip training will be optimal.  This methodology supposes that the hardware network is only used for inference, which is satisfactory for a wide range of applications. 
Online training leveraging the physics of the network would make it possible for the network to keep learning new tasks during its use. However, both the hyperparameter tuning and the sensitivity to errors make this goal challenging. More research is required to develop novel learning algorithms that are less sensitive than backpropagation to small variations of the parameters.

\section{\label{conclusion}Conclusion}
We have simulated a multi-layer neural network where the neurons are spintronic oscillators. Leveraging the transient dynamics of the neurons for memory and non-linearity, we have performed a time series classification task, sequential digits. We demonstrate that this multi-layer dynamical spintronic network can be trained with BPTT and standard machine learning tools, which is key for scaling to more complex tasks. Despite having a bounded internal variable, the spintronic network achieves $89.83~\%$ accuracy, as good as a standard recurrent neural network. These results demonstrate the potential of spintronic oscillators for time-series processing. 
We have devised guidelines for the design of such networks and their training: the relaxation time must be longer than the time-scale of the input and the cumulative drive must be close to one. Once built, a dynamical spintronic network can be trained to learn tasks over a 5-fold range of input time-scales. 
We have shown that we can sparsify the network down to $50~\%$ connectivity before training, further reducing the footprint and energy consumption of the system. For the sequential digits tasks and the considered architecture, we estimate that the network would consume about 40~pJ per image.
Despite the large interest for time-series processing with physical oscillators -- whether spintronic, photonic, resistive or CMOS-based -- there is little work on systematic scalable training methods for oscillators. Here, we show how to use standard machine learning tools to train multilayer spintronic oscillators, a method that is scalable. We also point on important design rules regarding the matching of the network dynamics with the input characteristics. Our use of a simplified model that is general enough makes our work easily adaptable to other types of physical oscillators. 
In this work we have demonstrated the concept with a simple task. Future studies will tackle larger networks and state of the art machine learning tasks.
These results open the path to the scaling up of spintronic neural networks for time-series processing at low energy cost.

\section*{Acknowledgements}
This work was supported by the European Union’s Horizon 2020 research and innovation programme under grant RadioSpin no. 101017098. 

\section*{Appendix}

\begin{figure}
    \centering
    \includegraphics[width=0.5\textwidth]{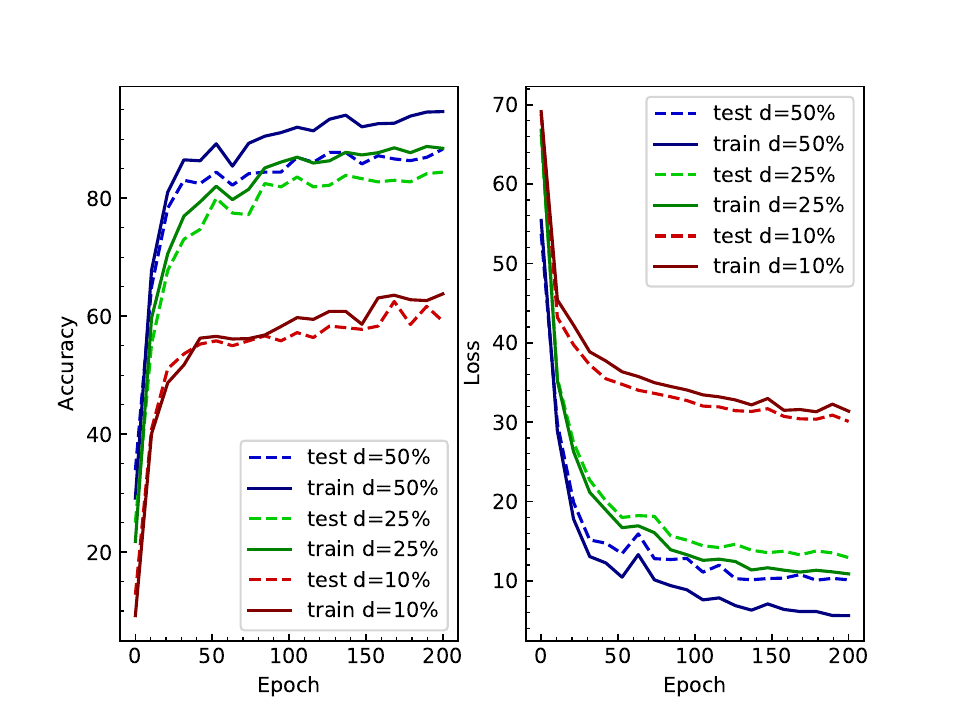}
    \caption{Right: Evolution of the accuracy (in \% of correctly classified inputs) versus the number of epochs. Left: Evolution of the loss versus the number of epochs. The train dataset is represented with solid lines and the test dataset with dashed lines. The colors blue, green and red correspond to global connection densities of 50\%, 25\% and 10\% respectively.}
    \label{fig:overfitting}
\end{figure}

We study the impact of the connection density on the overfitting.  Figure \ref{fig:overfitting} shows the evolution of the accuracy and loss versus the number of epochs, for the train and test datasets at different global connection densities. The conditions are the same as described in Section \ref{time series}.  We observe that the gap between the train and test datasets decreases with the connection density. This shows that reducing the density helps to prevent overfitting and explains why the optimal density is 50\% rather than 100\%.

\bibliography{biblio}

\begin{thebibliography}{44}%
\makeatletter
\providecommand \@ifxundefined [1]{%
 \@ifx{#1\undefined}
}%
\providecommand \@ifnum [1]{%
 \ifnum #1\expandafter \@firstoftwo
 \else \expandafter \@secondoftwo
 \fi
}%
\providecommand \@ifx [1]{%
 \ifx #1\expandafter \@firstoftwo
 \else \expandafter \@secondoftwo
 \fi
}%
\providecommand \natexlab [1]{#1}%
\providecommand \enquote  [1]{``#1''}%
\providecommand \bibnamefont  [1]{#1}%
\providecommand \bibfnamefont [1]{#1}%
\providecommand \citenamefont [1]{#1}%
\providecommand \href@noop [0]{\@secondoftwo}%
\providecommand \href [0]{\begingroup \@sanitize@url \@href}%
\providecommand \@href[1]{\@@startlink{#1}\@@href}%
\providecommand \@@href[1]{\endgroup#1\@@endlink}%
\providecommand \@sanitize@url [0]{\catcode `\\12\catcode `\$12\catcode `\&12\catcode `\#12\catcode `\^12\catcode `\_12\catcode `\%12\relax}%
\providecommand \@@startlink[1]{}%
\providecommand \@@endlink[0]{}%
\providecommand \url  [0]{\begingroup\@sanitize@url \@url }%
\providecommand \@url [1]{\endgroup\@href {#1}{\urlprefix }}%
\providecommand \urlprefix  [0]{URL }%
\providecommand \Eprint [0]{\href }%
\providecommand \doibase [0]{https://doi.org/}%
\providecommand \selectlanguage [0]{\@gobble}%
\providecommand \bibinfo  [0]{\@secondoftwo}%
\providecommand \bibfield  [0]{\@secondoftwo}%
\providecommand \translation [1]{[#1]}%
\providecommand \BibitemOpen [0]{}%
\providecommand \bibitemStop [0]{}%
\providecommand \bibitemNoStop [0]{.\EOS\space}%
\providecommand \EOS [0]{\spacefactor3000\relax}%
\providecommand \BibitemShut  [1]{\csname bibitem#1\endcsname}%
\let\auto@bib@innerbib\@empty
\bibitem [{\citenamefont {Pineda}(1987)}]{pineda1987generalization}%
  \BibitemOpen
  \bibfield  {author} {\bibinfo {author} {\bibfnamefont {F.}~\bibnamefont {Pineda}},\ }\bibfield  {title} {\bibinfo {title} {Generalization of back propagation to recurrent and higher order neural networks},\ }in\ \href@noop {} {\emph {\bibinfo {booktitle} {Neural information processing systems}}}\ (\bibinfo {year} {1987})\BibitemShut {NoStop}%
\bibitem [{\citenamefont {Lecun}(1992)}]{lecun1992theoretical}%
  \BibitemOpen
  \bibfield  {author} {\bibinfo {author} {\bibfnamefont {Y.}~\bibnamefont {Lecun}},\ }\bibfield  {title} {\bibinfo {title} {A theoretical framework for back-propagation},\ }in\ \href@noop {} {\emph {\bibinfo {booktitle} {Artificial Neural Networks: concepts and theory}}}\ (\bibinfo  {publisher} {IEEE Computer Society Press},\ \bibinfo {year} {1992})\BibitemShut {NoStop}%
\bibitem [{\citenamefont {Hermans}\ and\ \citenamefont {Schrauwen}(2013)}]{hermans2013training}%
  \BibitemOpen
  \bibfield  {author} {\bibinfo {author} {\bibfnamefont {M.}~\bibnamefont {Hermans}}\ and\ \bibinfo {author} {\bibfnamefont {B.}~\bibnamefont {Schrauwen}},\ }\bibfield  {title} {\bibinfo {title} {Training and analysing deep recurrent neural networks},\ }\href@noop {} {\bibfield  {journal} {\bibinfo  {journal} {Advances in neural information processing systems}\ }\textbf {\bibinfo {volume} {26}} (\bibinfo {year} {2013})}\BibitemShut {NoStop}%
\bibitem [{\citenamefont {Hasani}\ \emph {et~al.}(2021)\citenamefont {Hasani}, \citenamefont {Lechner}, \citenamefont {Amini}, \citenamefont {Rus},\ and\ \citenamefont {Grosu}}]{hasani2021liquid}%
  \BibitemOpen
  \bibfield  {author} {\bibinfo {author} {\bibfnamefont {R.}~\bibnamefont {Hasani}}, \bibinfo {author} {\bibfnamefont {M.}~\bibnamefont {Lechner}}, \bibinfo {author} {\bibfnamefont {A.}~\bibnamefont {Amini}}, \bibinfo {author} {\bibfnamefont {D.}~\bibnamefont {Rus}},\ and\ \bibinfo {author} {\bibfnamefont {R.}~\bibnamefont {Grosu}},\ }\bibfield  {title} {\bibinfo {title} {Liquid time-constant networks},\ }in\ \href@noop {} {\emph {\bibinfo {booktitle} {Proceedings of the AAAI Conference on Artificial Intelligence}}},\ Vol.~\bibinfo {volume} {35}\ (\bibinfo {year} {2021})\ pp.\ \bibinfo {pages} {7657--7666}\BibitemShut {NoStop}%
\bibitem [{\citenamefont {Eshraghian}\ \emph {et~al.}(2023)\citenamefont {Eshraghian}, \citenamefont {Ward}, \citenamefont {Neftci}, \citenamefont {Wang}, \citenamefont {Lenz}, \citenamefont {Dwivedi}, \citenamefont {Bennamoun}, \citenamefont {Jeong},\ and\ \citenamefont {Lu}}]{eshraghian2023training}%
  \BibitemOpen
  \bibfield  {author} {\bibinfo {author} {\bibfnamefont {J.~K.}\ \bibnamefont {Eshraghian}}, \bibinfo {author} {\bibfnamefont {M.}~\bibnamefont {Ward}}, \bibinfo {author} {\bibfnamefont {E.~O.}\ \bibnamefont {Neftci}}, \bibinfo {author} {\bibfnamefont {X.}~\bibnamefont {Wang}}, \bibinfo {author} {\bibfnamefont {G.}~\bibnamefont {Lenz}}, \bibinfo {author} {\bibfnamefont {G.}~\bibnamefont {Dwivedi}}, \bibinfo {author} {\bibfnamefont {M.}~\bibnamefont {Bennamoun}}, \bibinfo {author} {\bibfnamefont {D.~S.}\ \bibnamefont {Jeong}},\ and\ \bibinfo {author} {\bibfnamefont {W.~D.}\ \bibnamefont {Lu}},\ }\bibfield  {title} {\bibinfo {title} {Training spiking neural networks using lessons from deep learning},\ }\href@noop {} {\bibfield  {journal} {\bibinfo  {journal} {Proceedings of the IEEE}\ } (\bibinfo {year} {2023})}\BibitemShut {NoStop}%
\bibitem [{\citenamefont {Taherkhani}\ \emph {et~al.}(2020)\citenamefont {Taherkhani}, \citenamefont {Belatreche}, \citenamefont {Li}, \citenamefont {Cosma}, \citenamefont {Maguire},\ and\ \citenamefont {McGinnity}}]{taherkhani2020review}%
  \BibitemOpen
  \bibfield  {author} {\bibinfo {author} {\bibfnamefont {A.}~\bibnamefont {Taherkhani}}, \bibinfo {author} {\bibfnamefont {A.}~\bibnamefont {Belatreche}}, \bibinfo {author} {\bibfnamefont {Y.}~\bibnamefont {Li}}, \bibinfo {author} {\bibfnamefont {G.}~\bibnamefont {Cosma}}, \bibinfo {author} {\bibfnamefont {L.~P.}\ \bibnamefont {Maguire}},\ and\ \bibinfo {author} {\bibfnamefont {T.~M.}\ \bibnamefont {McGinnity}},\ }\bibfield  {title} {\bibinfo {title} {A review of learning in biologically plausible spiking neural networks},\ }\href@noop {} {\bibfield  {journal} {\bibinfo  {journal} {Neural Networks}\ }\textbf {\bibinfo {volume} {122}},\ \bibinfo {pages} {253} (\bibinfo {year} {2020})}\BibitemShut {NoStop}%
\bibitem [{\citenamefont {Tavanaei}\ \emph {et~al.}(2019)\citenamefont {Tavanaei}, \citenamefont {Ghodrati}, \citenamefont {Kheradpisheh}, \citenamefont {Masquelier},\ and\ \citenamefont {Maida}}]{tavanaei2019deep}%
  \BibitemOpen
  \bibfield  {author} {\bibinfo {author} {\bibfnamefont {A.}~\bibnamefont {Tavanaei}}, \bibinfo {author} {\bibfnamefont {M.}~\bibnamefont {Ghodrati}}, \bibinfo {author} {\bibfnamefont {S.~R.}\ \bibnamefont {Kheradpisheh}}, \bibinfo {author} {\bibfnamefont {T.}~\bibnamefont {Masquelier}},\ and\ \bibinfo {author} {\bibfnamefont {A.}~\bibnamefont {Maida}},\ }\bibfield  {title} {\bibinfo {title} {Deep learning in spiking neural networks},\ }\href@noop {} {\bibfield  {journal} {\bibinfo  {journal} {Neural networks}\ }\textbf {\bibinfo {volume} {111}},\ \bibinfo {pages} {47} (\bibinfo {year} {2019})}\BibitemShut {NoStop}%
\bibitem [{\citenamefont {Yin}\ \emph {et~al.}(2021)\citenamefont {Yin}, \citenamefont {Corradi},\ and\ \citenamefont {Boht{\'e}}}]{yin2021accurate}%
  \BibitemOpen
  \bibfield  {author} {\bibinfo {author} {\bibfnamefont {B.}~\bibnamefont {Yin}}, \bibinfo {author} {\bibfnamefont {F.}~\bibnamefont {Corradi}},\ and\ \bibinfo {author} {\bibfnamefont {S.~M.}\ \bibnamefont {Boht{\'e}}},\ }\bibfield  {title} {\bibinfo {title} {Accurate and efficient time-domain classification with adaptive spiking recurrent neural networks},\ }\href@noop {} {\bibfield  {journal} {\bibinfo  {journal} {Nature Machine Intelligence}\ }\textbf {\bibinfo {volume} {3}},\ \bibinfo {pages} {905} (\bibinfo {year} {2021})}\BibitemShut {NoStop}%
\bibitem [{\citenamefont {Chen}\ \emph {et~al.}(2018)\citenamefont {Chen}, \citenamefont {Rubanova}, \citenamefont {Bettencourt},\ and\ \citenamefont {Duvenaud}}]{NODE}%
  \BibitemOpen
  \bibfield  {author} {\bibinfo {author} {\bibfnamefont {R.~T.}\ \bibnamefont {Chen}}, \bibinfo {author} {\bibfnamefont {Y.}~\bibnamefont {Rubanova}}, \bibinfo {author} {\bibfnamefont {J.}~\bibnamefont {Bettencourt}},\ and\ \bibinfo {author} {\bibfnamefont {D.~K.}\ \bibnamefont {Duvenaud}},\ }\bibfield  {title} {\bibinfo {title} {Neural ordinary differential equations},\ }\href@noop {} {\bibfield  {journal} {\bibinfo  {journal} {Advances in neural information processing systems}\ }\textbf {\bibinfo {volume} {31}} (\bibinfo {year} {2018})}\BibitemShut {NoStop}%
\bibitem [{\citenamefont {Pehle}\ \emph {et~al.}(2022)\citenamefont {Pehle}, \citenamefont {Billaudelle}, \citenamefont {Cramer}, \citenamefont {Kaiser}, \citenamefont {Schreiber}, \citenamefont {Stradmann}, \citenamefont {Weis}, \citenamefont {Leibfried}, \citenamefont {M{\"u}ller},\ and\ \citenamefont {Schemmel}}]{pehle2022brainscales}%
  \BibitemOpen
  \bibfield  {author} {\bibinfo {author} {\bibfnamefont {C.}~\bibnamefont {Pehle}}, \bibinfo {author} {\bibfnamefont {S.}~\bibnamefont {Billaudelle}}, \bibinfo {author} {\bibfnamefont {B.}~\bibnamefont {Cramer}}, \bibinfo {author} {\bibfnamefont {J.}~\bibnamefont {Kaiser}}, \bibinfo {author} {\bibfnamefont {K.}~\bibnamefont {Schreiber}}, \bibinfo {author} {\bibfnamefont {Y.}~\bibnamefont {Stradmann}}, \bibinfo {author} {\bibfnamefont {J.}~\bibnamefont {Weis}}, \bibinfo {author} {\bibfnamefont {A.}~\bibnamefont {Leibfried}}, \bibinfo {author} {\bibfnamefont {E.}~\bibnamefont {M{\"u}ller}},\ and\ \bibinfo {author} {\bibfnamefont {J.}~\bibnamefont {Schemmel}},\ }\bibfield  {title} {\bibinfo {title} {The brainscales-2 accelerated neuromorphic system with hybrid plasticity},\ }\href@noop {} {\bibfield  {journal} {\bibinfo  {journal} {Frontiers in Neuroscience}\ }\textbf {\bibinfo {volume} {16}},\ \bibinfo {pages} {795876} (\bibinfo {year} {2022})}\BibitemShut {NoStop}%
\bibitem [{\citenamefont {Bueno}\ \emph {et~al.}(2018)\citenamefont {Bueno}, \citenamefont {Maktoobi}, \citenamefont {Froehly}, \citenamefont {Fischer}, \citenamefont {Jacquot}, \citenamefont {Larger},\ and\ \citenamefont {Brunner}}]{bueno2018reinforcement}%
  \BibitemOpen
  \bibfield  {author} {\bibinfo {author} {\bibfnamefont {J.}~\bibnamefont {Bueno}}, \bibinfo {author} {\bibfnamefont {S.}~\bibnamefont {Maktoobi}}, \bibinfo {author} {\bibfnamefont {L.}~\bibnamefont {Froehly}}, \bibinfo {author} {\bibfnamefont {I.}~\bibnamefont {Fischer}}, \bibinfo {author} {\bibfnamefont {M.}~\bibnamefont {Jacquot}}, \bibinfo {author} {\bibfnamefont {L.}~\bibnamefont {Larger}},\ and\ \bibinfo {author} {\bibfnamefont {D.}~\bibnamefont {Brunner}},\ }\bibfield  {title} {\bibinfo {title} {Reinforcement learning in a large-scale photonic recurrent neural network},\ }\href@noop {} {\bibfield  {journal} {\bibinfo  {journal} {Optica}\ }\textbf {\bibinfo {volume} {5}},\ \bibinfo {pages} {756} (\bibinfo {year} {2018})}\BibitemShut {NoStop}%
\bibitem [{\citenamefont {Qu}\ \emph {et~al.}(2022)\citenamefont {Qu}, \citenamefont {Zhou}, \citenamefont {Khoram}, \citenamefont {Yu},\ and\ \citenamefont {Yu}}]{qu2022resonance}%
  \BibitemOpen
  \bibfield  {author} {\bibinfo {author} {\bibfnamefont {Y.}~\bibnamefont {Qu}}, \bibinfo {author} {\bibfnamefont {M.}~\bibnamefont {Zhou}}, \bibinfo {author} {\bibfnamefont {E.}~\bibnamefont {Khoram}}, \bibinfo {author} {\bibfnamefont {N.}~\bibnamefont {Yu}},\ and\ \bibinfo {author} {\bibfnamefont {Z.}~\bibnamefont {Yu}},\ }\bibfield  {title} {\bibinfo {title} {Resonance for analog recurrent neural network},\ }\href@noop {} {\bibfield  {journal} {\bibinfo  {journal} {ACS Photonics}\ }\textbf {\bibinfo {volume} {9}},\ \bibinfo {pages} {1647} (\bibinfo {year} {2022})}\BibitemShut {NoStop}%
\bibitem [{\citenamefont {Coulombe}\ \emph {et~al.}(2017)\citenamefont {Coulombe}, \citenamefont {York},\ and\ \citenamefont {Sylvestre}}]{coulombe2017computing}%
  \BibitemOpen
  \bibfield  {author} {\bibinfo {author} {\bibfnamefont {J.~C.}\ \bibnamefont {Coulombe}}, \bibinfo {author} {\bibfnamefont {M.~C.}\ \bibnamefont {York}},\ and\ \bibinfo {author} {\bibfnamefont {J.}~\bibnamefont {Sylvestre}},\ }\bibfield  {title} {\bibinfo {title} {Computing with networks of nonlinear mechanical oscillators},\ }\href@noop {} {\bibfield  {journal} {\bibinfo  {journal} {PloS one}\ }\textbf {\bibinfo {volume} {12}},\ \bibinfo {pages} {e0178663} (\bibinfo {year} {2017})}\BibitemShut {NoStop}%
\bibitem [{\citenamefont {Hughes}\ \emph {et~al.}(2019)\citenamefont {Hughes}, \citenamefont {Williamson}, \citenamefont {Minkov},\ and\ \citenamefont {Fan}}]{hughes2019wave}%
  \BibitemOpen
  \bibfield  {author} {\bibinfo {author} {\bibfnamefont {T.~W.}\ \bibnamefont {Hughes}}, \bibinfo {author} {\bibfnamefont {I.~A.}\ \bibnamefont {Williamson}}, \bibinfo {author} {\bibfnamefont {M.}~\bibnamefont {Minkov}},\ and\ \bibinfo {author} {\bibfnamefont {S.}~\bibnamefont {Fan}},\ }\bibfield  {title} {\bibinfo {title} {Wave physics as an analog recurrent neural network},\ }\href@noop {} {\bibfield  {journal} {\bibinfo  {journal} {Science advances}\ }\textbf {\bibinfo {volume} {5}},\ \bibinfo {pages} {eaay6946} (\bibinfo {year} {2019})}\BibitemShut {NoStop}%
\bibitem [{\citenamefont {Finocchio}\ \emph {et~al.}(2021)\citenamefont {Finocchio}, \citenamefont {Di~Ventra}, \citenamefont {Camsari}, \citenamefont {Everschor-Sitte}, \citenamefont {Amiri},\ and\ \citenamefont {Zeng}}]{finocchio2021promise}%
  \BibitemOpen
  \bibfield  {author} {\bibinfo {author} {\bibfnamefont {G.}~\bibnamefont {Finocchio}}, \bibinfo {author} {\bibfnamefont {M.}~\bibnamefont {Di~Ventra}}, \bibinfo {author} {\bibfnamefont {K.~Y.}\ \bibnamefont {Camsari}}, \bibinfo {author} {\bibfnamefont {K.}~\bibnamefont {Everschor-Sitte}}, \bibinfo {author} {\bibfnamefont {P.~K.}\ \bibnamefont {Amiri}},\ and\ \bibinfo {author} {\bibfnamefont {Z.}~\bibnamefont {Zeng}},\ }\bibfield  {title} {\bibinfo {title} {The promise of spintronics for unconventional computing},\ }\href@noop {} {\bibfield  {journal} {\bibinfo  {journal} {Journal of Magnetism and Magnetic Materials}\ }\textbf {\bibinfo {volume} {521}},\ \bibinfo {pages} {167506} (\bibinfo {year} {2021})}\BibitemShut {NoStop}%
\bibitem [{\citenamefont {Grollier}\ \emph {et~al.}(2020)\citenamefont {Grollier}, \citenamefont {Querlioz}, \citenamefont {Camsari}, \citenamefont {Everschor-Sitte}, \citenamefont {Fukami},\ and\ \citenamefont {Stiles}}]{grollier2020neuromorphic}%
  \BibitemOpen
  \bibfield  {author} {\bibinfo {author} {\bibfnamefont {J.}~\bibnamefont {Grollier}}, \bibinfo {author} {\bibfnamefont {D.}~\bibnamefont {Querlioz}}, \bibinfo {author} {\bibfnamefont {K.}~\bibnamefont {Camsari}}, \bibinfo {author} {\bibfnamefont {K.}~\bibnamefont {Everschor-Sitte}}, \bibinfo {author} {\bibfnamefont {S.}~\bibnamefont {Fukami}},\ and\ \bibinfo {author} {\bibfnamefont {M.~D.}\ \bibnamefont {Stiles}},\ }\bibfield  {title} {\bibinfo {title} {Neuromorphic spintronics},\ }\href@noop {} {\bibfield  {journal} {\bibinfo  {journal} {Nature electronics}\ }\textbf {\bibinfo {volume} {3}},\ \bibinfo {pages} {360} (\bibinfo {year} {2020})}\BibitemShut {NoStop}%
\bibitem [{\citenamefont {Tanaka}\ \emph {et~al.}(2019)\citenamefont {Tanaka}, \citenamefont {Yamane}, \citenamefont {H{\'e}roux}, \citenamefont {Nakane}, \citenamefont {Kanazawa}, \citenamefont {Takeda}, \citenamefont {Numata}, \citenamefont {Nakano},\ and\ \citenamefont {Hirose}}]{tanaka2019recent}%
  \BibitemOpen
  \bibfield  {author} {\bibinfo {author} {\bibfnamefont {G.}~\bibnamefont {Tanaka}}, \bibinfo {author} {\bibfnamefont {T.}~\bibnamefont {Yamane}}, \bibinfo {author} {\bibfnamefont {J.~B.}\ \bibnamefont {H{\'e}roux}}, \bibinfo {author} {\bibfnamefont {R.}~\bibnamefont {Nakane}}, \bibinfo {author} {\bibfnamefont {N.}~\bibnamefont {Kanazawa}}, \bibinfo {author} {\bibfnamefont {S.}~\bibnamefont {Takeda}}, \bibinfo {author} {\bibfnamefont {H.}~\bibnamefont {Numata}}, \bibinfo {author} {\bibfnamefont {D.}~\bibnamefont {Nakano}},\ and\ \bibinfo {author} {\bibfnamefont {A.}~\bibnamefont {Hirose}},\ }\bibfield  {title} {\bibinfo {title} {Recent advances in physical reservoir computing: A review},\ }\href@noop {} {\bibfield  {journal} {\bibinfo  {journal} {Neural Networks}\ }\textbf {\bibinfo {volume} {115}},\ \bibinfo {pages} {100} (\bibinfo {year} {2019})}\BibitemShut {NoStop}%
\bibitem [{\citenamefont {Torrejon}\ \emph {et~al.}(2017)\citenamefont {Torrejon}, \citenamefont {Riou}, \citenamefont {Araujo}, \citenamefont {Tsunegi}, \citenamefont {Khalsa}, \citenamefont {Querlioz}, \citenamefont {Bortolotti}, \citenamefont {Cros}, \citenamefont {Yakushiji}, \citenamefont {Fukushima} \emph {et~al.}}]{torrejon2017neuromorphic}%
  \BibitemOpen
  \bibfield  {author} {\bibinfo {author} {\bibfnamefont {J.}~\bibnamefont {Torrejon}}, \bibinfo {author} {\bibfnamefont {M.}~\bibnamefont {Riou}}, \bibinfo {author} {\bibfnamefont {F.~A.}\ \bibnamefont {Araujo}}, \bibinfo {author} {\bibfnamefont {S.}~\bibnamefont {Tsunegi}}, \bibinfo {author} {\bibfnamefont {G.}~\bibnamefont {Khalsa}}, \bibinfo {author} {\bibfnamefont {D.}~\bibnamefont {Querlioz}}, \bibinfo {author} {\bibfnamefont {P.}~\bibnamefont {Bortolotti}}, \bibinfo {author} {\bibfnamefont {V.}~\bibnamefont {Cros}}, \bibinfo {author} {\bibfnamefont {K.}~\bibnamefont {Yakushiji}}, \bibinfo {author} {\bibfnamefont {A.}~\bibnamefont {Fukushima}}, \emph {et~al.},\ }\bibfield  {title} {\bibinfo {title} {Neuromorphic computing with nanoscale spintronic oscillators},\ }\href@noop {} {\bibfield  {journal} {\bibinfo  {journal} {Nature}\ }\textbf {\bibinfo {volume} {547}},\ \bibinfo {pages} {428} (\bibinfo {year} {2017})}\BibitemShut {NoStop}%
\bibitem [{\citenamefont {Markovic}\ \emph {et~al.}(2019)\citenamefont {Markovic}, \citenamefont {Leroux}, \citenamefont {Riou}, \citenamefont {Abreu~Araujo}, \citenamefont {Torrejon}, \citenamefont {Querlioz}, \citenamefont {Fukushima}, \citenamefont {Yuasa}, \citenamefont {Trastoy}, \citenamefont {Bortolotti} \emph {et~al.}}]{markovic2019reservoir}%
  \BibitemOpen
  \bibfield  {author} {\bibinfo {author} {\bibfnamefont {D.}~\bibnamefont {Markovic}}, \bibinfo {author} {\bibfnamefont {N.}~\bibnamefont {Leroux}}, \bibinfo {author} {\bibfnamefont {M.}~\bibnamefont {Riou}}, \bibinfo {author} {\bibfnamefont {F.}~\bibnamefont {Abreu~Araujo}}, \bibinfo {author} {\bibfnamefont {J.}~\bibnamefont {Torrejon}}, \bibinfo {author} {\bibfnamefont {D.}~\bibnamefont {Querlioz}}, \bibinfo {author} {\bibfnamefont {A.}~\bibnamefont {Fukushima}}, \bibinfo {author} {\bibfnamefont {S.}~\bibnamefont {Yuasa}}, \bibinfo {author} {\bibfnamefont {J.}~\bibnamefont {Trastoy}}, \bibinfo {author} {\bibfnamefont {P.}~\bibnamefont {Bortolotti}}, \emph {et~al.},\ }\bibfield  {title} {\bibinfo {title} {Reservoir computing with the frequency, phase, and amplitude of spin-torque nano-oscillators},\ }\href@noop {} {\bibfield  {journal} {\bibinfo  {journal} {Applied Physics Letters}\ }\textbf {\bibinfo {volume} {114}} (\bibinfo {year} {2019})}\BibitemShut {NoStop}%
\bibitem [{\citenamefont {Tsunegi}\ \emph {et~al.}(2019)\citenamefont {Tsunegi}, \citenamefont {Taniguchi}, \citenamefont {Nakajima}, \citenamefont {Miwa}, \citenamefont {Yakushiji}, \citenamefont {Fukushima}, \citenamefont {Yuasa},\ and\ \citenamefont {Kubota}}]{tsunegi2019physical}%
  \BibitemOpen
  \bibfield  {author} {\bibinfo {author} {\bibfnamefont {S.}~\bibnamefont {Tsunegi}}, \bibinfo {author} {\bibfnamefont {T.}~\bibnamefont {Taniguchi}}, \bibinfo {author} {\bibfnamefont {K.}~\bibnamefont {Nakajima}}, \bibinfo {author} {\bibfnamefont {S.}~\bibnamefont {Miwa}}, \bibinfo {author} {\bibfnamefont {K.}~\bibnamefont {Yakushiji}}, \bibinfo {author} {\bibfnamefont {A.}~\bibnamefont {Fukushima}}, \bibinfo {author} {\bibfnamefont {S.}~\bibnamefont {Yuasa}},\ and\ \bibinfo {author} {\bibfnamefont {H.}~\bibnamefont {Kubota}},\ }\bibfield  {title} {\bibinfo {title} {Physical reservoir computing based on spin torque oscillator with forced synchronization},\ }\href@noop {} {\bibfield  {journal} {\bibinfo  {journal} {Applied Physics Letters}\ }\textbf {\bibinfo {volume} {114}} (\bibinfo {year} {2019})}\BibitemShut {NoStop}%
\bibitem [{\citenamefont {Kanao}\ \emph {et~al.}(2019)\citenamefont {Kanao}, \citenamefont {Suto}, \citenamefont {Mizushima}, \citenamefont {Goto}, \citenamefont {Tanamoto},\ and\ \citenamefont {Nagasawa}}]{kanao2019reservoir}%
  \BibitemOpen
  \bibfield  {author} {\bibinfo {author} {\bibfnamefont {T.}~\bibnamefont {Kanao}}, \bibinfo {author} {\bibfnamefont {H.}~\bibnamefont {Suto}}, \bibinfo {author} {\bibfnamefont {K.}~\bibnamefont {Mizushima}}, \bibinfo {author} {\bibfnamefont {H.}~\bibnamefont {Goto}}, \bibinfo {author} {\bibfnamefont {T.}~\bibnamefont {Tanamoto}},\ and\ \bibinfo {author} {\bibfnamefont {T.}~\bibnamefont {Nagasawa}},\ }\bibfield  {title} {\bibinfo {title} {Reservoir computing on spin-torque oscillator array},\ }\href@noop {} {\bibfield  {journal} {\bibinfo  {journal} {Physical Review Applied}\ }\textbf {\bibinfo {volume} {12}},\ \bibinfo {pages} {024052} (\bibinfo {year} {2019})}\BibitemShut {NoStop}%
\bibitem [{\citenamefont {Shreya}\ \emph {et~al.}(2023)\citenamefont {Shreya}, \citenamefont {Jenkins}, \citenamefont {Rezaeiyan}, \citenamefont {Li}, \citenamefont {B{\"o}hnert}, \citenamefont {Benetti}, \citenamefont {Ferreira}, \citenamefont {Moradi},\ and\ \citenamefont {Farkhani}}]{shreya2023granular}%
  \BibitemOpen
  \bibfield  {author} {\bibinfo {author} {\bibfnamefont {S.}~\bibnamefont {Shreya}}, \bibinfo {author} {\bibfnamefont {A.}~\bibnamefont {Jenkins}}, \bibinfo {author} {\bibfnamefont {Y.}~\bibnamefont {Rezaeiyan}}, \bibinfo {author} {\bibfnamefont {R.}~\bibnamefont {Li}}, \bibinfo {author} {\bibfnamefont {T.}~\bibnamefont {B{\"o}hnert}}, \bibinfo {author} {\bibfnamefont {L.}~\bibnamefont {Benetti}}, \bibinfo {author} {\bibfnamefont {R.}~\bibnamefont {Ferreira}}, \bibinfo {author} {\bibfnamefont {F.}~\bibnamefont {Moradi}},\ and\ \bibinfo {author} {\bibfnamefont {H.}~\bibnamefont {Farkhani}},\ }\bibfield  {title} {\bibinfo {title} {Granular vortex spin-torque nano oscillator for reservoir computing},\ }\href@noop {} {\bibfield  {journal} {\bibinfo  {journal} {Scientific Reports}\ }\textbf {\bibinfo {volume} {13}},\ \bibinfo {pages} {16722} (\bibinfo {year} {2023})}\BibitemShut {NoStop}%
\bibitem [{\citenamefont {Tsunegi}\ \emph {et~al.}(2018)\citenamefont {Tsunegi}, \citenamefont {Taniguchi}, \citenamefont {Miwa}, \citenamefont {Nakajima}, \citenamefont {Yakushiji}, \citenamefont {Fukushima}, \citenamefont {Yuasa},\ and\ \citenamefont {Kubota}}]{tsunegi2018evaluation}%
  \BibitemOpen
  \bibfield  {author} {\bibinfo {author} {\bibfnamefont {S.}~\bibnamefont {Tsunegi}}, \bibinfo {author} {\bibfnamefont {T.}~\bibnamefont {Taniguchi}}, \bibinfo {author} {\bibfnamefont {S.}~\bibnamefont {Miwa}}, \bibinfo {author} {\bibfnamefont {K.}~\bibnamefont {Nakajima}}, \bibinfo {author} {\bibfnamefont {K.}~\bibnamefont {Yakushiji}}, \bibinfo {author} {\bibfnamefont {A.}~\bibnamefont {Fukushima}}, \bibinfo {author} {\bibfnamefont {S.}~\bibnamefont {Yuasa}},\ and\ \bibinfo {author} {\bibfnamefont {H.}~\bibnamefont {Kubota}},\ }\bibfield  {title} {\bibinfo {title} {Evaluation of memory capacity of spin torque oscillator for recurrent neural networks},\ }\href@noop {} {\bibfield  {journal} {\bibinfo  {journal} {Japanese Journal of Applied Physics}\ }\textbf {\bibinfo {volume} {57}},\ \bibinfo {pages} {120307} (\bibinfo {year} {2018})}\BibitemShut {NoStop}%
\bibitem [{\citenamefont {Romera}\ \emph {et~al.}(2018)\citenamefont {Romera}, \citenamefont {Talatchian}, \citenamefont {Tsunegi}, \citenamefont {Abreu~Araujo}, \citenamefont {Cros}, \citenamefont {Bortolotti}, \citenamefont {Trastoy}, \citenamefont {Yakushiji}, \citenamefont {Fukushima}, \citenamefont {Kubota} \emph {et~al.}}]{romera2018vowel}%
  \BibitemOpen
  \bibfield  {author} {\bibinfo {author} {\bibfnamefont {M.}~\bibnamefont {Romera}}, \bibinfo {author} {\bibfnamefont {P.}~\bibnamefont {Talatchian}}, \bibinfo {author} {\bibfnamefont {S.}~\bibnamefont {Tsunegi}}, \bibinfo {author} {\bibfnamefont {F.}~\bibnamefont {Abreu~Araujo}}, \bibinfo {author} {\bibfnamefont {V.}~\bibnamefont {Cros}}, \bibinfo {author} {\bibfnamefont {P.}~\bibnamefont {Bortolotti}}, \bibinfo {author} {\bibfnamefont {J.}~\bibnamefont {Trastoy}}, \bibinfo {author} {\bibfnamefont {K.}~\bibnamefont {Yakushiji}}, \bibinfo {author} {\bibfnamefont {A.}~\bibnamefont {Fukushima}}, \bibinfo {author} {\bibfnamefont {H.}~\bibnamefont {Kubota}}, \emph {et~al.},\ }\bibfield  {title} {\bibinfo {title} {Vowel recognition with four coupled spin-torque nano-oscillators},\ }\href@noop {} {\bibfield  {journal} {\bibinfo  {journal} {Nature}\ }\textbf {\bibinfo {volume} {563}},\ \bibinfo {pages} {230} (\bibinfo {year} {2018})}\BibitemShut {NoStop}%
\bibitem [{\citenamefont {Zahedinejad}\ \emph {et~al.}(2020)\citenamefont {Zahedinejad}, \citenamefont {Awad}, \citenamefont {Muralidhar}, \citenamefont {Khymyn}, \citenamefont {Fulara}, \citenamefont {Mazraati}, \citenamefont {Dvornik},\ and\ \citenamefont {{\AA}kerman}}]{zahedinejad2020two}%
  \BibitemOpen
  \bibfield  {author} {\bibinfo {author} {\bibfnamefont {M.}~\bibnamefont {Zahedinejad}}, \bibinfo {author} {\bibfnamefont {A.~A.}\ \bibnamefont {Awad}}, \bibinfo {author} {\bibfnamefont {S.}~\bibnamefont {Muralidhar}}, \bibinfo {author} {\bibfnamefont {R.}~\bibnamefont {Khymyn}}, \bibinfo {author} {\bibfnamefont {H.}~\bibnamefont {Fulara}}, \bibinfo {author} {\bibfnamefont {H.}~\bibnamefont {Mazraati}}, \bibinfo {author} {\bibfnamefont {M.}~\bibnamefont {Dvornik}},\ and\ \bibinfo {author} {\bibfnamefont {J.}~\bibnamefont {{\AA}kerman}},\ }\bibfield  {title} {\bibinfo {title} {Two-dimensional mutually synchronized spin hall nano-oscillator arrays for neuromorphic computing},\ }\href@noop {} {\bibfield  {journal} {\bibinfo  {journal} {Nature nanotechnology}\ }\textbf {\bibinfo {volume} {15}},\ \bibinfo {pages} {47} (\bibinfo {year} {2020})}\BibitemShut {NoStop}%
\bibitem [{\citenamefont {Ross}\ \emph {et~al.}(2023)\citenamefont {Ross}, \citenamefont {Leroux}, \citenamefont {De~Riz}, \citenamefont {Markovi{\'c}}, \citenamefont {Sanz-Hern{\'a}ndez}, \citenamefont {Trastoy}, \citenamefont {Bortolotti}, \citenamefont {Querlioz}, \citenamefont {Martins}, \citenamefont {Benetti} \emph {et~al.}}]{Andrewmulti}%
  \BibitemOpen
  \bibfield  {author} {\bibinfo {author} {\bibfnamefont {A.}~\bibnamefont {Ross}}, \bibinfo {author} {\bibfnamefont {N.}~\bibnamefont {Leroux}}, \bibinfo {author} {\bibfnamefont {A.}~\bibnamefont {De~Riz}}, \bibinfo {author} {\bibfnamefont {D.}~\bibnamefont {Markovi{\'c}}}, \bibinfo {author} {\bibfnamefont {D.}~\bibnamefont {Sanz-Hern{\'a}ndez}}, \bibinfo {author} {\bibfnamefont {J.}~\bibnamefont {Trastoy}}, \bibinfo {author} {\bibfnamefont {P.}~\bibnamefont {Bortolotti}}, \bibinfo {author} {\bibfnamefont {D.}~\bibnamefont {Querlioz}}, \bibinfo {author} {\bibfnamefont {L.}~\bibnamefont {Martins}}, \bibinfo {author} {\bibfnamefont {L.}~\bibnamefont {Benetti}}, \emph {et~al.},\ }\bibfield  {title} {\bibinfo {title} {Multilayer spintronic neural networks with radiofrequency connections},\ }\href@noop {} {\bibfield  {journal} {\bibinfo  {journal} {Nature Nanotechnology}\ }\textbf {\bibinfo {volume} {18}},\ \bibinfo {pages} {1273} (\bibinfo {year} {2023})}\BibitemShut {NoStop}%
\bibitem [{\citenamefont {Rodrigues}\ \emph {et~al.}(2023)\citenamefont {Rodrigues}, \citenamefont {Raimondo}, \citenamefont {Puliafito}, \citenamefont {Moukhader}, \citenamefont {Azzerboni}, \citenamefont {Hamadeh}, \citenamefont {Pirro}, \citenamefont {Carpentieri},\ and\ \citenamefont {Finocchio}}]{DynFinnochio}%
  \BibitemOpen
  \bibfield  {author} {\bibinfo {author} {\bibfnamefont {D.}~\bibnamefont {Rodrigues}}, \bibinfo {author} {\bibfnamefont {E.}~\bibnamefont {Raimondo}}, \bibinfo {author} {\bibfnamefont {V.}~\bibnamefont {Puliafito}}, \bibinfo {author} {\bibfnamefont {R.}~\bibnamefont {Moukhader}}, \bibinfo {author} {\bibfnamefont {B.}~\bibnamefont {Azzerboni}}, \bibinfo {author} {\bibfnamefont {A.}~\bibnamefont {Hamadeh}}, \bibinfo {author} {\bibfnamefont {P.}~\bibnamefont {Pirro}}, \bibinfo {author} {\bibfnamefont {M.}~\bibnamefont {Carpentieri}},\ and\ \bibinfo {author} {\bibfnamefont {G.}~\bibnamefont {Finocchio}},\ }\bibfield  {title} {\bibinfo {title} {Dynamical neural network based on spin transfer nano-oscillators},\ }\href {https://doi.org/10.1109/TNANO.2023.3330535} {\bibfield  {journal} {\bibinfo  {journal} {IEEE Transactions on Nanotechnology}\ }\textbf {\bibinfo {volume} {PP}},\ \bibinfo {pages} {1} (\bibinfo {year} {2023})}\BibitemShut {NoStop}%
\bibitem [{\citenamefont {Slavin}\ and\ \citenamefont {Tiberkevich}(2009)}]{slavin_nonlinear_2009}%
  \BibitemOpen
  \bibfield  {author} {\bibinfo {author} {\bibfnamefont {A.}~\bibnamefont {Slavin}}\ and\ \bibinfo {author} {\bibfnamefont {V.}~\bibnamefont {Tiberkevich}},\ }\bibfield  {title} {\bibinfo {title} {Nonlinear {Auto}-{Oscillator} {Theory} of {Microwave} {Generation} by {Spin}-{Polarized} {Current}},\ }\href {https://doi.org/10.1109/TMAG.2008.2009935} {\bibfield  {journal} {\bibinfo  {journal} {IEEE Transactions on Magnetics}\ }\textbf {\bibinfo {volume} {45}},\ \bibinfo {pages} {1875} (\bibinfo {year} {2009})}\BibitemShut {NoStop}%
\bibitem [{\citenamefont {Pedregosa}\ \emph {et~al.}(2011)\citenamefont {Pedregosa}, \citenamefont {Varoquaux}, \citenamefont {Gramfort}, \citenamefont {Michel}, \citenamefont {Thirion}, \citenamefont {Grisel}, \citenamefont {Blondel}, \citenamefont {Prettenhofer}, \citenamefont {Weiss}, \citenamefont {Dubourg}, \citenamefont {Vanderplas}, \citenamefont {Passos}, \citenamefont {Cournapeau}, \citenamefont {Brucher}, \citenamefont {Perrot},\ and\ \citenamefont {Duchesnay}}]{scikit-learn}%
  \BibitemOpen
  \bibfield  {author} {\bibinfo {author} {\bibfnamefont {F.}~\bibnamefont {Pedregosa}}, \bibinfo {author} {\bibfnamefont {G.}~\bibnamefont {Varoquaux}}, \bibinfo {author} {\bibfnamefont {A.}~\bibnamefont {Gramfort}}, \bibinfo {author} {\bibfnamefont {V.}~\bibnamefont {Michel}}, \bibinfo {author} {\bibfnamefont {B.}~\bibnamefont {Thirion}}, \bibinfo {author} {\bibfnamefont {O.}~\bibnamefont {Grisel}}, \bibinfo {author} {\bibfnamefont {M.}~\bibnamefont {Blondel}}, \bibinfo {author} {\bibfnamefont {P.}~\bibnamefont {Prettenhofer}}, \bibinfo {author} {\bibfnamefont {R.}~\bibnamefont {Weiss}}, \bibinfo {author} {\bibfnamefont {V.}~\bibnamefont {Dubourg}}, \bibinfo {author} {\bibfnamefont {J.}~\bibnamefont {Vanderplas}}, \bibinfo {author} {\bibfnamefont {A.}~\bibnamefont {Passos}}, \bibinfo {author} {\bibfnamefont {D.}~\bibnamefont {Cournapeau}}, \bibinfo {author} {\bibfnamefont {M.}~\bibnamefont {Brucher}}, \bibinfo {author} {\bibfnamefont {M.}~\bibnamefont {Perrot}},\ and\ \bibinfo {author} {\bibfnamefont
  {E.}~\bibnamefont {Duchesnay}},\ }\bibfield  {title} {\bibinfo {title} {Scikit-learn: Machine learning in {P}ython},\ }\href@noop {} {\bibfield  {journal} {\bibinfo  {journal} {Journal of Machine Learning Research}\ }\textbf {\bibinfo {volume} {12}},\ \bibinfo {pages} {2825} (\bibinfo {year} {2011})}\BibitemShut {NoStop}%
\bibitem [{\citenamefont {Robinson}\ and\ \citenamefont {Fallside}(1987)}]{robinson1987utilityBPTT}%
  \BibitemOpen
  \bibfield  {author} {\bibinfo {author} {\bibfnamefont {A.~J.}\ \bibnamefont {Robinson}}\ and\ \bibinfo {author} {\bibfnamefont {F.}~\bibnamefont {Fallside}},\ }\href@noop {} {\emph {\bibinfo {title} {The utility driven dynamic error propagation network}}},\ Vol.~\bibinfo {volume} {11}\ (\bibinfo  {publisher} {University of Cambridge Department of Engineering Cambridge},\ \bibinfo {year} {1987})\BibitemShut {NoStop}%
\bibitem [{\citenamefont {Werbos}(1988)}]{werbos1988generalization}%
  \BibitemOpen
  \bibfield  {author} {\bibinfo {author} {\bibfnamefont {P.~J.}\ \bibnamefont {Werbos}},\ }\bibfield  {title} {\bibinfo {title} {Generalization of backpropagation with application to a recurrent gas market model},\ }\href@noop {} {\bibfield  {journal} {\bibinfo  {journal} {Neural networks}\ }\textbf {\bibinfo {volume} {1}},\ \bibinfo {pages} {339} (\bibinfo {year} {1988})}\BibitemShut {NoStop}%
\bibitem [{\citenamefont {Mozer}(2013)}]{mozer2013focused}%
  \BibitemOpen
  \bibfield  {author} {\bibinfo {author} {\bibfnamefont {M.~C.}\ \bibnamefont {Mozer}},\ }\bibfield  {title} {\bibinfo {title} {A focused backpropagation algorithm for temporal pattern recognition},\ }in\ \href@noop {} {\emph {\bibinfo {booktitle} {Backpropagation}}}\ (\bibinfo  {publisher} {Psychology Press},\ \bibinfo {year} {2013})\ pp.\ \bibinfo {pages} {137--169}\BibitemShut {NoStop}%
\bibitem [{\citenamefont {Akiba}\ \emph {et~al.}(2019)\citenamefont {Akiba}, \citenamefont {Sano}, \citenamefont {Yanase}, \citenamefont {Ohta},\ and\ \citenamefont {Koyama}}]{optuna_2019}%
  \BibitemOpen
  \bibfield  {author} {\bibinfo {author} {\bibfnamefont {T.}~\bibnamefont {Akiba}}, \bibinfo {author} {\bibfnamefont {S.}~\bibnamefont {Sano}}, \bibinfo {author} {\bibfnamefont {T.}~\bibnamefont {Yanase}}, \bibinfo {author} {\bibfnamefont {T.}~\bibnamefont {Ohta}},\ and\ \bibinfo {author} {\bibfnamefont {M.}~\bibnamefont {Koyama}},\ }\bibfield  {title} {\bibinfo {title} {Optuna: A next-generation hyperparameter optimization framework},\ }in\ \href@noop {} {\emph {\bibinfo {booktitle} {Proceedings of the 25th {ACM} {SIGKDD} International Conference on Knowledge Discovery and Data Mining}}}\ (\bibinfo {year} {2019})\BibitemShut {NoStop}%
\bibitem [{\citenamefont {Kingma}\ and\ \citenamefont {Ba}(2017)}]{kingma2017adammethodstochasticoptimization}%
  \BibitemOpen
  \bibfield  {author} {\bibinfo {author} {\bibfnamefont {D.~P.}\ \bibnamefont {Kingma}}\ and\ \bibinfo {author} {\bibfnamefont {J.}~\bibnamefont {Ba}},\ }\href {https://arxiv.org/abs/1412.6980} {\bibinfo {title} {Adam: A method for stochastic optimization}} (\bibinfo {year} {2017}),\ \Eprint {https://arxiv.org/abs/1412.6980} {arXiv:1412.6980 [cs.LG]} \BibitemShut {NoStop}%
\bibitem [{\citenamefont {Choi}\ \emph {et~al.}(2022)\citenamefont {Choi}, \citenamefont {Park}, \citenamefont {Kang}, \citenamefont {Kim}, \citenamefont {Rieh}, \citenamefont {Lee}, \citenamefont {Kim},\ and\ \citenamefont {Park}}]{choi2022voltage}%
  \BibitemOpen
  \bibfield  {author} {\bibinfo {author} {\bibfnamefont {J.-G.}\ \bibnamefont {Choi}}, \bibinfo {author} {\bibfnamefont {J.}~\bibnamefont {Park}}, \bibinfo {author} {\bibfnamefont {M.-G.}\ \bibnamefont {Kang}}, \bibinfo {author} {\bibfnamefont {D.}~\bibnamefont {Kim}}, \bibinfo {author} {\bibfnamefont {J.-S.}\ \bibnamefont {Rieh}}, \bibinfo {author} {\bibfnamefont {K.-J.}\ \bibnamefont {Lee}}, \bibinfo {author} {\bibfnamefont {K.-J.}\ \bibnamefont {Kim}},\ and\ \bibinfo {author} {\bibfnamefont {B.-G.}\ \bibnamefont {Park}},\ }\bibfield  {title} {\bibinfo {title} {Voltage-driven gigahertz frequency tuning of spin hall nano-oscillators},\ }\href@noop {} {\bibfield  {journal} {\bibinfo  {journal} {Nature communications}\ }\textbf {\bibinfo {volume} {13}},\ \bibinfo {pages} {3783} (\bibinfo {year} {2022})}\BibitemShut {NoStop}%
\bibitem [{\citenamefont {Funahashi}\ and\ \citenamefont {Nakamura}(1993)}]{funahashi1993approximation}%
  \BibitemOpen
  \bibfield  {author} {\bibinfo {author} {\bibfnamefont {K.-i.}\ \bibnamefont {Funahashi}}\ and\ \bibinfo {author} {\bibfnamefont {Y.}~\bibnamefont {Nakamura}},\ }\bibfield  {title} {\bibinfo {title} {Approximation of dynamical systems by continuous time recurrent neural networks},\ }\href@noop {} {\bibfield  {journal} {\bibinfo  {journal} {Neural networks}\ }\textbf {\bibinfo {volume} {6}},\ \bibinfo {pages} {801} (\bibinfo {year} {1993})}\BibitemShut {NoStop}%
\bibitem [{\citenamefont {Borders}\ \emph {et~al.}(2024)\citenamefont {Borders}, \citenamefont {Madhavan}, \citenamefont {Daniels}, \citenamefont {Georgiou}, \citenamefont {Lueker-Boden}, \citenamefont {Santos}, \citenamefont {Braganca}, \citenamefont {Stiles}, \citenamefont {McClelland},\ and\ \citenamefont {Hoskins}}]{borders2024measurement}%
  \BibitemOpen
  \bibfield  {author} {\bibinfo {author} {\bibfnamefont {W.~A.}\ \bibnamefont {Borders}}, \bibinfo {author} {\bibfnamefont {A.}~\bibnamefont {Madhavan}}, \bibinfo {author} {\bibfnamefont {M.~W.}\ \bibnamefont {Daniels}}, \bibinfo {author} {\bibfnamefont {V.}~\bibnamefont {Georgiou}}, \bibinfo {author} {\bibfnamefont {M.}~\bibnamefont {Lueker-Boden}}, \bibinfo {author} {\bibfnamefont {T.~S.}\ \bibnamefont {Santos}}, \bibinfo {author} {\bibfnamefont {P.~M.}\ \bibnamefont {Braganca}}, \bibinfo {author} {\bibfnamefont {M.~D.}\ \bibnamefont {Stiles}}, \bibinfo {author} {\bibfnamefont {J.~J.}\ \bibnamefont {McClelland}},\ and\ \bibinfo {author} {\bibfnamefont {B.~D.}\ \bibnamefont {Hoskins}},\ }\bibfield  {title} {\bibinfo {title} {Measurement-driven neural-network training for integrated magnetic tunnel junction arrays},\ }\href@noop {} {\bibfield  {journal} {\bibinfo  {journal} {Physical Review Applied}\ }\textbf {\bibinfo {volume} {21}},\ \bibinfo {pages} {054028} (\bibinfo {year} {2024})}\BibitemShut {NoStop}%
\bibitem [{\citenamefont {Jung}\ \emph {et~al.}(2022)\citenamefont {Jung}, \citenamefont {Lee}, \citenamefont {Myung}, \citenamefont {Kim}, \citenamefont {Yoon}, \citenamefont {Kwon}, \citenamefont {Ju}, \citenamefont {Kim}, \citenamefont {Yi}, \citenamefont {Han} \emph {et~al.}}]{jung2022crossbar}%
  \BibitemOpen
  \bibfield  {author} {\bibinfo {author} {\bibfnamefont {S.}~\bibnamefont {Jung}}, \bibinfo {author} {\bibfnamefont {H.}~\bibnamefont {Lee}}, \bibinfo {author} {\bibfnamefont {S.}~\bibnamefont {Myung}}, \bibinfo {author} {\bibfnamefont {H.}~\bibnamefont {Kim}}, \bibinfo {author} {\bibfnamefont {S.~K.}\ \bibnamefont {Yoon}}, \bibinfo {author} {\bibfnamefont {S.-W.}\ \bibnamefont {Kwon}}, \bibinfo {author} {\bibfnamefont {Y.}~\bibnamefont {Ju}}, \bibinfo {author} {\bibfnamefont {M.}~\bibnamefont {Kim}}, \bibinfo {author} {\bibfnamefont {W.}~\bibnamefont {Yi}}, \bibinfo {author} {\bibfnamefont {S.}~\bibnamefont {Han}}, \emph {et~al.},\ }\bibfield  {title} {\bibinfo {title} {A crossbar array of magnetoresistive memory devices for in-memory computing},\ }\href@noop {} {\bibfield  {journal} {\bibinfo  {journal} {Nature}\ }\textbf {\bibinfo {volume} {601}},\ \bibinfo {pages} {211} (\bibinfo {year} {2022})}\BibitemShut {NoStop}%
\bibitem [{\citenamefont {Nikonov}\ \emph {et~al.}(2020)\citenamefont {Nikonov}, \citenamefont {Kurahashi}, \citenamefont {Ayers}, \citenamefont {Li}, \citenamefont {Kamgaing}, \citenamefont {Dogiamis}, \citenamefont {Lee}, \citenamefont {Fan},\ and\ \citenamefont {Young}}]{nikonov2020convolution}%
  \BibitemOpen
  \bibfield  {author} {\bibinfo {author} {\bibfnamefont {D.~E.}\ \bibnamefont {Nikonov}}, \bibinfo {author} {\bibfnamefont {P.}~\bibnamefont {Kurahashi}}, \bibinfo {author} {\bibfnamefont {J.~S.}\ \bibnamefont {Ayers}}, \bibinfo {author} {\bibfnamefont {H.}~\bibnamefont {Li}}, \bibinfo {author} {\bibfnamefont {T.}~\bibnamefont {Kamgaing}}, \bibinfo {author} {\bibfnamefont {G.~C.}\ \bibnamefont {Dogiamis}}, \bibinfo {author} {\bibfnamefont {H.-J.}\ \bibnamefont {Lee}}, \bibinfo {author} {\bibfnamefont {Y.}~\bibnamefont {Fan}},\ and\ \bibinfo {author} {\bibfnamefont {I.}~\bibnamefont {Young}},\ }\bibfield  {title} {\bibinfo {title} {Convolution inference via synchronization of a coupled cmos oscillator array},\ }\href@noop {} {\bibfield  {journal} {\bibinfo  {journal} {IEEE Journal on Exploratory Solid-State Computational Devices and Circuits}\ }\textbf {\bibinfo {volume} {6}},\ \bibinfo {pages} {170} (\bibinfo {year} {2020})}\BibitemShut {NoStop}%
\bibitem [{\citenamefont {Nath}\ \emph {et~al.}(2023)\citenamefont {Nath}, \citenamefont {Sun}, \citenamefont {Nandi}, \citenamefont {Chen}, \citenamefont {Wang}, \citenamefont {Das}, \citenamefont {Lei}, \citenamefont {Faraone}, \citenamefont {Rickard},\ and\ \citenamefont {Elliman}}]{nath2023harnessing}%
  \BibitemOpen
  \bibfield  {author} {\bibinfo {author} {\bibfnamefont {S.~K.}\ \bibnamefont {Nath}}, \bibinfo {author} {\bibfnamefont {X.}~\bibnamefont {Sun}}, \bibinfo {author} {\bibfnamefont {S.~K.}\ \bibnamefont {Nandi}}, \bibinfo {author} {\bibfnamefont {X.}~\bibnamefont {Chen}}, \bibinfo {author} {\bibfnamefont {Z.}~\bibnamefont {Wang}}, \bibinfo {author} {\bibfnamefont {S.~K.}\ \bibnamefont {Das}}, \bibinfo {author} {\bibfnamefont {W.}~\bibnamefont {Lei}}, \bibinfo {author} {\bibfnamefont {L.}~\bibnamefont {Faraone}}, \bibinfo {author} {\bibfnamefont {W.~D.}\ \bibnamefont {Rickard}},\ and\ \bibinfo {author} {\bibfnamefont {R.~G.}\ \bibnamefont {Elliman}},\ }\bibfield  {title} {\bibinfo {title} {Harnessing metal/oxide interlayer to engineer the memristive response and oscillation dynamics of two-terminal memristors},\ }\href@noop {} {\bibfield  {journal} {\bibinfo  {journal} {Advanced Functional Materials}\ }\textbf {\bibinfo {volume} {33}},\ \bibinfo {pages} {2306428} (\bibinfo {year} {2023})}\BibitemShut {NoStop}%
\bibitem [{\citenamefont {Feldmann}\ \emph {et~al.}(2019)\citenamefont {Feldmann}, \citenamefont {Youngblood}, \citenamefont {Wright}, \citenamefont {Bhaskaran},\ and\ \citenamefont {Pernice}}]{feldmann2019all}%
  \BibitemOpen
  \bibfield  {author} {\bibinfo {author} {\bibfnamefont {J.}~\bibnamefont {Feldmann}}, \bibinfo {author} {\bibfnamefont {N.}~\bibnamefont {Youngblood}}, \bibinfo {author} {\bibfnamefont {C.~D.}\ \bibnamefont {Wright}}, \bibinfo {author} {\bibfnamefont {H.}~\bibnamefont {Bhaskaran}},\ and\ \bibinfo {author} {\bibfnamefont {W.~H.}\ \bibnamefont {Pernice}},\ }\bibfield  {title} {\bibinfo {title} {All-optical spiking neurosynaptic networks with self-learning capabilities},\ }\href@noop {} {\bibfield  {journal} {\bibinfo  {journal} {Nature}\ }\textbf {\bibinfo {volume} {569}},\ \bibinfo {pages} {208} (\bibinfo {year} {2019})}\BibitemShut {NoStop}%
\bibitem [{\citenamefont {Wright}\ \emph {et~al.}(2022)\citenamefont {Wright}, \citenamefont {Onodera}, \citenamefont {Stein}, \citenamefont {Wang}, \citenamefont {Schachter}, \citenamefont {Hu},\ and\ \citenamefont {McMahon}}]{wright2022deep}%
  \BibitemOpen
  \bibfield  {author} {\bibinfo {author} {\bibfnamefont {L.~G.}\ \bibnamefont {Wright}}, \bibinfo {author} {\bibfnamefont {T.}~\bibnamefont {Onodera}}, \bibinfo {author} {\bibfnamefont {M.~M.}\ \bibnamefont {Stein}}, \bibinfo {author} {\bibfnamefont {T.}~\bibnamefont {Wang}}, \bibinfo {author} {\bibfnamefont {D.~T.}\ \bibnamefont {Schachter}}, \bibinfo {author} {\bibfnamefont {Z.}~\bibnamefont {Hu}},\ and\ \bibinfo {author} {\bibfnamefont {P.~L.}\ \bibnamefont {McMahon}},\ }\bibfield  {title} {\bibinfo {title} {Deep physical neural networks trained with backpropagation},\ }\href@noop {} {\bibfield  {journal} {\bibinfo  {journal} {Nature}\ }\textbf {\bibinfo {volume} {601}},\ \bibinfo {pages} {549} (\bibinfo {year} {2022})}\BibitemShut {NoStop}%
\bibitem [{\citenamefont {Rasch}\ \emph {et~al.}(2023)\citenamefont {Rasch}, \citenamefont {Mackin}, \citenamefont {Le~Gallo}, \citenamefont {Chen}, \citenamefont {Fasoli}, \citenamefont {Odermatt}, \citenamefont {Li}, \citenamefont {Nandakumar}, \citenamefont {Narayanan}, \citenamefont {Tsai} \emph {et~al.}}]{rasch2023hardware}%
  \BibitemOpen
  \bibfield  {author} {\bibinfo {author} {\bibfnamefont {M.~J.}\ \bibnamefont {Rasch}}, \bibinfo {author} {\bibfnamefont {C.}~\bibnamefont {Mackin}}, \bibinfo {author} {\bibfnamefont {M.}~\bibnamefont {Le~Gallo}}, \bibinfo {author} {\bibfnamefont {A.}~\bibnamefont {Chen}}, \bibinfo {author} {\bibfnamefont {A.}~\bibnamefont {Fasoli}}, \bibinfo {author} {\bibfnamefont {F.}~\bibnamefont {Odermatt}}, \bibinfo {author} {\bibfnamefont {N.}~\bibnamefont {Li}}, \bibinfo {author} {\bibfnamefont {S.}~\bibnamefont {Nandakumar}}, \bibinfo {author} {\bibfnamefont {P.}~\bibnamefont {Narayanan}}, \bibinfo {author} {\bibfnamefont {H.}~\bibnamefont {Tsai}}, \emph {et~al.},\ }\bibfield  {title} {\bibinfo {title} {Hardware-aware training for large-scale and diverse deep learning inference workloads using in-memory computing-based accelerators},\ }\href@noop {} {\bibfield  {journal} {\bibinfo  {journal} {Nature communications}\ }\textbf {\bibinfo {volume} {14}},\ \bibinfo {pages} {5282} (\bibinfo {year} {2023})}\BibitemShut
  {NoStop}%
\bibitem [{\citenamefont {Xu}\ \emph {et~al.}(2024)\citenamefont {Xu}, \citenamefont {Luo}, \citenamefont {Liu}, \citenamefont {Fan}, \citenamefont {Xiao}, \citenamefont {Wang}, \citenamefont {Wang},\ and\ \citenamefont {Huang}}]{xu2024perfecting}%
  \BibitemOpen
  \bibfield  {author} {\bibinfo {author} {\bibfnamefont {T.}~\bibnamefont {Xu}}, \bibinfo {author} {\bibfnamefont {Z.}~\bibnamefont {Luo}}, \bibinfo {author} {\bibfnamefont {S.}~\bibnamefont {Liu}}, \bibinfo {author} {\bibfnamefont {L.}~\bibnamefont {Fan}}, \bibinfo {author} {\bibfnamefont {Q.}~\bibnamefont {Xiao}}, \bibinfo {author} {\bibfnamefont {B.}~\bibnamefont {Wang}}, \bibinfo {author} {\bibfnamefont {D.}~\bibnamefont {Wang}},\ and\ \bibinfo {author} {\bibfnamefont {C.}~\bibnamefont {Huang}},\ }\bibfield  {title} {\bibinfo {title} {Perfecting imperfect physical neural networks with transferable robustness using sharpness-aware training},\ }\href@noop {} {\bibfield  {journal} {\bibinfo  {journal} {arXiv preprint arXiv:2411.12352}\ } (\bibinfo {year} {2024})}\BibitemShut {NoStop}%
\end{thebibliography}%

\end{document}